\documentstyle[12pt]{article}

\begin{document}
\newlength{\bas}
\setlength{\bas}{\baselineskip}

\title{Analysis of data \\
on low energy $\pi N \rightarrow \pi \pi N$ reaction. \\
I. Total cross sections.
\protect {\footnote{This work was funded by the
Deutsche Forschungsgemeinschaft (DFG),
Bundesminister f\"ur Forschung und Technologie (BMFT), and the
Kernforschungsanlage J\"ulich (KFA) and Russian Ministry
of Education Grant G13.018}}}

\author{
V.V. Vereshagin${}^a$,\ S.G. Sherman${}^b$,\ A.N.Manashov${}^a$, \\
U. Bohnert${}^c$,\  M. Dillig${}^d$,\  W. Eyrich${}^c$,\
O. J\"akel${}^d$,\ M. Moosburger${}^c$  \\ \\
${}^a)$ St. Petersburg State University, St. Petersburg, Russia\\  \\
${}^b)$ St. Petersburg Nuclear Physics Insitute,
St. Petersburg, Russia\\ \\
${}^c)$ Physikalisches Institut IV der Universit\"at Erlangen-N\"urnberg,\\
Erlangen, Germany \\  \\
${}^d) $Institut f\"ur Theoretische Physik III der Universit\"at
Erlangen-N\"urnberg,\\
Erlangen, Germany \\ \\    }

\date{\today}
\begin{titlepage}
\maketitle

\begin{abstract}
This is the first of a series of papers on a consistent
model
independent analysis of the complete experimental information on
the reaction $\pi N \rightarrow \pi \pi N$
at pion momenta up to 500 MeV/c.
The paper summarizes the theoretical approach
and details of the computational procedure. The complete database  on
total cross sections in 5 $\pi \pi N$ channels is given
together with a critical discussion of their model independent analysis.
\renewcommand{\baselinestretch}{0.5}
\end{abstract}
\end{titlepage}

\sloppy

\section {Introduction}
\setlength{\baselineskip}{\bas}

The
$\pi N \rightarrow \pi \pi N$
processes near threshold is currently a subject of intensive
investigations, both theoretically and experimentally [1--25].
There are two main reasons why this process has
attracted growing interest: a) the development of
experimental techniques made it  possible to  measure
in the nearest neighborhood of threshold not only total
cross sections
but also angular distributions; b) reliable  data on the
parameters of low energy $\pi \pi$ interaction are necessary
for a precise estimate of higher order corrections in Chiral
Perturbation Theory (ChPT), which gives a link between QCD and
the world of hadrons (see, for example, \cite{26}).
The reaction
in question provides an excellent opportunity to get the
parameters appearing in  the ChPT expansions \cite{27,28,29}.

In view of the importance of reliable and sufficiently precise results
(which has been specially stressed
in a recent paper
\cite{30},
devoted to the problem of estimating
of higher order terms of ChPT)
it is desirable to perform
a complete, model independent analysis
based on the
full set of available experimental information
on
$\pi N \rightarrow \pi \pi N$ reaction
near threshold.
The main goal of such analysis would be to establish the
phenomenological form of the reaction amplitude
in the low energy domain and to fix (with the help of data) the
values of the corresponding parameters appearing in this form.

It is clear that this rather complex task is
preferentially solved in various steps,
as the systematic  experimental and, especially,
theoretical investigation of the low energy dynamics of the
reaction in question is, in fact, just beginning. The
problem of the extraction of low energy $\pi\pi$ -- interaction
parameters with a desirable accuracy looks, in such a situation,
extremely difficult. At the same time, the successive  resolution of
this problem is important by its own, irrespective of the
following application of the results: the pion, which is the
lightest hadron, participates --- really or virtually --- in
every strong interaction process (see, for example,
\cite{28,31,32}). This is  a reason why the
complete phenomenological analysis of the data on low energy
$\pi N \rightarrow \pi \pi N$ reaction
is really of actual interest.

In the present paper we make a first step in this direction.
It combines two points:
it includes the phenomenological ansatz for the amplitude
(the generalization of the standard approach used in the
isotopic analysis) and the corresponding analysis of the full
set of experimental data on total cross sections of 5 channels
at $P_{lab}\leq 500 MeV/c$. The goal of this paper is to answer
the following questions:
\begin{enumerate}
\item
Is the modern database on total cross sections internally
consistent or are there in a statistical sense ``doubtful''
points?
\item
How strong is the comparative influence of the various
mechanisms ($\pi$ -- and $\Delta$-- exchanges, ``background'') on
the
$\pi N \rightarrow \pi \pi N$ reaction
amplitude?
\item
What is the capability of data to distinguish between the
different contributions?
\item
Is it possible to extract the values of the
$\pi N \rightarrow \pi \pi N$
amplitude parameters from the data on total cross sections only
(of particular interest being the values of the parameters
describing the low energy $\pi\pi$--interaction)?
\end{enumerate}

The new precise data on $\pi N\rightarrow\pi\pi N$ total cross sections, which
became available in the course of the last years (see Table~1)
support the common expectation to contain accurate
information on the low energy $\pi\pi$ interaction near
threshold. As pointed out above, this information would be of
great importance in connection with the problem of a
quantitative description of the low energy hadron dynamics from
QCD.

To extract the information on $\pi\pi$ -- scattering from the
data on
$\pi N\rightarrow\pi\pi N$
reactions, one has to perform a refined analysis,
because the detailed mechanism of the latter reaction is
not understood in detail and the influence of the $4\pi$--vertex being
not necessarily
dominant. Its contribution could be  (if at all) separated
among all others with the help of an accurate analysis of the
full set of experimental data including those on
differential cross sections.

At this moment, however, the overwhelming majority of recent
results on $\pi\pi$--parameters is based on analyses of
data on total cross sections only. Most of the analyses having
been performed with the help of the Olsson--Turner (OT) model
\cite{64}  (for a review see \cite{20}). This situation
is similar to that  of the seventies (see, for example,
refs.~\cite{27,28}). That time the most popular method of the
corresponding data analysis was the Chew--Low extrapolation
procedure \cite{65}. This method, however, proved to be not
sufficiently reliable, the main uncertainty
originating from the phenomenological features of the
$\pi N\rightarrow\pi\pi N$
reaction: the contribution of the OPE graph turned out to be
small in comparison with  those of the other graphs (see, for
example, the recent papers \cite{61,62,63}). Due to this reason
the results of the extrapolation of experimental data to the
unphysical point $\tau=\mu^2$ become extremely sensitive to
small variations (errors) of the  data used. From
the purely mathematical point of view this effect is
just a
manifestation of J.\,Hadamards general theorem concerning
the ambiguities of extrapolation
problems (see ref.~\cite{33}).

The same situation repeats now again. A good example is given by
the comparison of the results on $S$--wave
$\pi\pi$--scattering lengths obtained by different groups, where
the most interesting conclusion has been made by the OMICRON
group in ref.~\cite{8}: a single parameter $\xi$ of the OT
model is not sufficient for the adequate description of the low
energy data on total cross sections of three channels
($\pi^-\pi^+ n$, $\pi^-\pi^0 p$,  $\pi^+\pi^+ n$). It should be
stressed here that from the purely technical point of view
the analysis made by this group is  one of the most accurate,
because it accounts for all specific features of the OT model. It
looks much as if the origin of the difficulty mentioned above
is hidden either in these very specific features of the OT model or in the
inconsistency of the data.

The original formulae given in ref.~\cite{64} are valid only
for the threshold values of the amplitudes. Thus, to extract
the value of $\xi$ (or, equivalently, $S$--wave scattering lengths)
one has to perform the extrapolation of the experimental data
on total cross sections to the threshold in all 5
channels of the
$\pi N\rightarrow\pi\pi N$
reaction. On this way, however, one meets at least two serious
obstacles closely connected to each other. First: to reduce
errors one must use for the extrapolation the datapoints
nearest to  threshold. Second: one needs to know
the  form of the extrapolation law. As it is well known the low
energy measurements exhibit extremely delicate experimental
problems by itself, the difficulties increase
approaching to threshold. Thus, those very points which
would be the most suitable for the extrapolation, are the
most difficult to measure.
It must be mentioned that  ---
in addition to purely experimental difficulties --- there is
the unresolved theoretical problem of data interpretation: in
the near threshold domain the radiative corrections become
significant. Furthermore, the form of the extrapolation law is more
or less known for the case of  elastic scattering  only (due to
general requirements given by quantum mechanics,
such as partial wave expansion, unitarity, etc.). In the case under
consideration --- the
$\pi N\rightarrow\pi\pi N$
reaction --- these limitations are considerably weaker and the
extrapolation law itself is still unknown.
%becomes the subject of investigation.

All these obstacles result in large
systematic errors of the parameters to be determined.
Besides, needless to mention, the OT model --- as
every other one --- has its own limits of applicability.

%We did not cleary understand the meaning of the following
%phrase.
Out of this situation we started a
general analysis of the phenomenological features of
the total cross sections of the 5 $\pi N\rightarrow\pi\pi N$
channels in the low energy domain ($p_{lab}\leq 500 MeV/c$). The main
goal of this analysis is to get a deeper understanding of the
phenomenology of the process and to estimate the informative capacity
and selfconsistency of the modern database.

This paper is the first  one  of a series devoted to the
systematic model independent analysis of \underline{all}
available experimental information on the low energy
$\pi N \rightarrow \pi \pi N$
reaction. It contains --- along with the analysis of  data on
total cross sections --- the detailed description of the
theoretical approach and the computational procedure, as both
subjects are important for the subsequent analysis of
distributions also.  That is why the text is divided into 4
sections.
The details of the approach are discussed in Sec.2, where we
present also the complete database on total cross sections.
Sec.3 contains the results of the theoretical analysis
of the experimental situation and those of the data fitting .
The discussion of the results obtained and the concluding
statements are given in Sec.4.

\section{Theoretical approach}
\subsection{Decomposition of the amplitude}

The main ingredients of our approach are given in refs.
{}~\cite{4,5,19}.
 Here, we just want to recall some important
points and give an extension of the formalism.

Our ansatz is model independent in the sense that we do not
make
a priori suggestions on the form of interactions
(other than the well
known symmetry requirements), coupling constants
(except the $\pi NN$ coupling) or on
the off--shell behaviour of the
%$\pi NN$ and $\pi N \Delta$
different vertices.
It is based on the exact symmetry principles
(Lorentz--invariance, isotopic and crossing symmetry,
C--, P-- and T--invariance
and Bose--statistics for pions) and some well established
features of the low
energy phenomenology, the basic idea being precisely
what is used for the
solution of a so called "incorrect" tasks
(frequently also called "ill--posed" or "ill--conditioned";
for a detailed review
of this subject see  ref. \cite{33} and refs.  quoted
therein).
Our procedure can be schematically formulated in the
following way:

1. Write down the amplitude $M$ of the process under
consideration in the form:
\begin{equation}
M = S + B,
\label{MMgl1}
\end{equation}
where $S$ stands for the ``special'' contributions with a strong
 dependence on kinematical variables,
 which must be separated and where $B$ includes all
other contributions (in accordance with the terminology of
ref.~\cite{19},
we call $B$ the "background").

2. Take into account all available information on the form of
the functions
$S$ and $B$. In our case, this step
implies the selection of convenient  parameterizations for
the above mentioned
functions in terms of the kinematical variables suitable
for our problem.
Thus, one obtains the amplitude $M$
of the reaction in question as a function
of some set of free parameters.
{}From the decomposition of the invariant
amplitude $M$ --- as motivated by the underlying
dynamics --- into $S$, with a
sensitive dependence on variables
(as will be discussed below, $S$
contains the OPE graph and the
$\Delta$--dominated amplitudes),
and a smooth background $B$, it is clear that the
parameterization of $S$ and $B$ differs in its functional
structure.

3. Fit experimental data with the amplitude $M$,
obtained in step 2,
to determine the influence of the various elements of
$M$ on the data and to find, if possible, the values of the
parameters or at least some combinations of them.

It is clear, that the more detailed information on
the form of the functions
$S$ and $B$ is involved, the stronger is the model dependence
of the results obtained.  Hence it is suitable here to fix
the meaning of the term "model independent
approach". To do this, we give here in a shortened form the
necessary formulae  (detailed explanations are given in ref.
\cite{19}).

The amplitude $M^{abc}_{fl,\mu\nu}$ of the reaction
\begin{equation}
\pi_a(k_1) + N_l^{\nu}(p) \rightarrow
     \pi_b(k_2) + \pi_c(k_3) + N_f^{\mu}(q)
\label{piN}
\end{equation}
($a,b,c=1,2,3$ and $f,l=1,2$ are isotopic indices;
$\mu,\nu=1,2$ are the nucleon
polarizations), can be written in the "canonical" form:
\begin{equation}
M^{abc}_{fl,\mu\nu} = (\tau^a)_{fl} \delta^{bc} A^{\mu\nu} +
(\tau^b)_{fl} \delta^{ac} B^{\mu\nu}+(\tau^c)_{fl}
     \delta^{ab} C^{\mu\nu}+
i\epsilon^{abc} \delta_{fl} D^{\mu\nu},
\label{Amp}
\end{equation}
where ($X=A,B,C,D$)
\begin{eqnarray}
X^{\mu\nu}& =&  \bar{u}^{\mu}(q) \hat{X} u^{\nu}(p),\label{X1} \\
\hat{X}& =& \{S_x \cdot I +
     \bar{V}_x \cdot \hat{k} + V_x \cdot \hat{\bar{k}}
+ \frac{i}{2} T_x [\hat{k},\hat{\bar{k}}] \} (i\gamma_5),
     \label{X2}
\\
k=-k_1+\epsilon k_2 + \bar{\epsilon} k_3,&\qquad&
\bar{k}=-k_1+\bar{\epsilon} k_2 +\epsilon k_3,
     \label{X3}
\\
\epsilon=exp(i \frac{2\pi}{3}),
     & \qquad& \bar{\epsilon}=\epsilon^*,
\label{X4}
\end{eqnarray}

\noindent
and $\hat{K}=\gamma_{\mu} K^{\mu}$ with $K=k,\bar{k}$.
The 16 invariant form factors $S_x,\bar{V}_x,V_x,T_x$ (with
$X=A,B,C,D$) are functions of five independent scalar
variables:

\begin{eqnarray}
\tau=(p-q)^2, \qquad  &\nu = (p+q)\cdot k, \qquad &
\theta=(p-q)\cdot k, \nonumber \\ &
\bar{\nu}=(p+q)\cdot\bar{k}, \qquad &
\bar{\theta}=(p-q)\cdot\bar{k}.
\label{Tau}
\end{eqnarray}

The general properties of these functions with respect
to the requirements of
Bose, crossing and C--symmetry are described in
\cite{19} and references therein.

So, up to this point, no suggestions have been made on the form
of the amplitude $M$ --- only the restrictions due to the exact
symmetries of strong interactions are being taken into account.

To proceed further, however,
one has to include some additional physical information.
First of all, it is necessary to choose the energy domain,
which is the most suitable for the solution of our task ---
the determination
of $\pi \pi$--scattering parameters.

In a previous paper \cite{19}
we have discussed only general features of our method.
The particular form of the parameterization
considered there is
suitable only for a sufficiently narrow interval of the incident beam
momentum $k_1$. In such a case the possible influence of
resonances (which were implied as being far from the
boundaries of the physical phase space) can be described by
slowly varying real functions.

 Thus, in a narrow momentum
 range, physical arguments suggest a representation of
 $M$ as a sum
of two terms:
1. the contribution of the OPE graph (denoted hereafter as
$\Pi$); 2. the total contribution of all graphs besides OPE
("background" denoted as $B$).  The crucial point is to note
that, inside of this narrow kinematical interval, the first
term (OPE) is characterized by a strong $\tau$ dependence
(due to the pole which is placed close to the boundary of the
physical phase space), while the second one ("background") has
no such strong dependence on the variables because of the
absence of the corresponding physical mechanisms. Reversely,
this means that the background contribution can be approximated
by any smooth function of the variables from eq. (\ref{Tau}).
The choice of
the approximating functions is certainly not unique (this
weakens
the meaning of the term "model independent approach").

In this note --- compared to \cite{19} ---
our goal is more ambitious;
we would like to extend our analysis
to the full kinematical region from
threshold (or some MeV above it,
to avoid problems caused by the rescattering
of final particles; see, for example, \cite{27}) up to
$k_{1,lab} \sim 0.5~GeV/c$.
 In this domain the
 $\pi N\rightarrow \pi\pi N$ reaction amplitude
is strongly influenced by the $\Delta$--isobar creation in
the initial ($k_{1,lab} \sim 0.3~GeV/c$)  and final
($k_{1,lab} \sim 0.5~GeV/c$) $\pi N$ states
\cite{19}.

In this case the contribution of the graphs with $\Delta$
isobar exchange (simply called $\Delta$ below) shows a
sensitive dependence on the scattering energy  due to its
resonant structure in the initial and final $\pi N$ invariant
masses. Therefore, it has to be separated from the
background amplitude.
This contribution cannot be approximated
by linear functions in the kinematical variables with the
desirable accuracy.  Furthermore, the behavior of the amplitude
near the resonance position is strongly restricted by
unitarity.  Consequently, our amplitude
has to be extended: we must add the
 $\Delta$--isobar contribution explicitly,
for which, in particular, the presence of an imaginary part is
necessary.

Keeping in mind all these notes above,
we arrive at the conclusion that the
appropriate way to construct
a suitable approximation for the amplitude M ---
in the momentum range investigated ---
is the decomposition:
\begin{equation}
M = B+\Pi + \Delta
\label{MMgl2}
\end{equation}
(i.e. $S=\Pi + \Delta$, eq. (\ref{MMgl1}), see also fig. 1),
with the following
amplitudes:

\begin{itemize}

\item[--]
B is the total contribution of all graphs besides those with
$\Delta$ poles and OPE; It  can be approximated by linear
polynomials with unknown coefficients.
This procedure gives 11
free parameters:
$A_1$ -- $A_{11}$,
(see Sec.2.2.1 below).

\item[--]
$\Pi$ is the main ingredient of the picture --- the OPE graph.
It
contains four free parameters: $g_0$ -- $g_3$,
(see Sec.2.2.2 below).

\item[--]
$\Delta$ is the $\Delta$--isobar contribution which can be approximated
with functions of the Breit--Wigner type (with the terms
$iM_{\Delta} \Gamma_{\Delta}$ in the denominator taken into
account).  The simplest way to write down the unknown
polynomials in the corresponding numerators is to use
effective Lagrangians, this method guaranteing the validity of
all necessary symmetry conditions (isotopic and discrete
symmetries, etc.).  According to Sec. 2.2.3 (see below)
this introduces 6
extra free parameters.

\end{itemize}

Taking everything into account, one obtains the amplitude
$M$ of the reaction (1) in the form given by the formulae
(\ref{Amp}--\ref{X4}),
with the functions $S_x,\bar{V}_x,V_x,T_x$ ($X=A,B,C,D$)
fixed by 21 free parameters to be determined with the help of
experimental data at
$k_{1,lab} \leq 500~MeV/c$.

We have to remember that only 10 of these 21 parameters
(namely, those, describing the OPE  and the
$\Delta$ contributions) can be interpreted as coupling
constants of the corresponding vertices,
while the 11 remaining in the background amplitude
have no clear relation to underlying physical parameters
(e.g. the $N^*$(1440)
position and width).

Some of 10 coupling constants mentioned above could be
fixed (or, at least, constrained) by additinal theoretical and
experimental information. For example, the $\pi N \Delta$
coupling could be fixed by the known value of the isobar decay
widths (or, equivalently, by the additive quark model) while
the $\Delta \pi \Delta$ one -- by the requirements of $SU(6)$
symmetry. We, howeveer, prefer -- for the sake of flexibility
of our procedure -- to keep them as free parameters. This
allows us to avoid the introducing of unnecessary model
dependence into our approach.

Of course, it is not to be expected that the restricted data
set on total cross sections only is able to fix all 21
parameters in quantitative way. To solve this task one has to
analyze  the full set of distributions at several values of
incident momentum. Our present goal is rather to answer the
questions listed in Sec.1 and to give insight on degree of
model dependence of the previously known results.

In closing this section we remark that in order to explore the
significance of the various pieces of $M$, both fits with
$M=B,\Pi,\Delta$ only and with $M=\Pi+B,\Delta+B,\Pi+\Delta$
were performed in comparison to the fit with the full amplitude
$M=B+\Pi+\Delta$.

\subsection{The amplitude}

The explicit inclusion of the $\Delta$ contribution is the main
extension of the formulae given in ref. \cite{19}. Therefore we
give in the following a detailed description of the formalism;
for completeness, we supplement this presentation by
the baswic ingredients for the parametrization of the OPE
and the background terms from ref.~\cite{19}.

\subsubsection{The background contribution}

As shown in \cite{19} the aggregate contribution of the backgroud
mechanisms can be described as follows:

\begin{eqnarray}
S_A&=&A_1+A_2\tau+A_3(\theta+\bar \theta) \nonumber \\
V_A&=&A_4+A_5\tau+A_6\theta+A_7 \bar\theta \\
\bar V_A&=&A_4+A_5\tau+A_6\bar \theta+A_7 \theta \nonumber \\
T_A&=&\imath A_8(\nu-\bar\nu)\nonumber
\end{eqnarray}

\begin{eqnarray}
S_B &=&A_1+A_2\tau+A_3(\bar \epsilon\theta+\epsilon\bar \theta) \nonumber \\
V_B&=&A_4+\epsilon A_5\tau+A_6\theta+\bar \epsilon A_7 \bar\theta \\
\bar V_B&=&\bar \epsilon A_4+\bar \epsilon A_5\tau+A_6\bar \theta+ \epsilon
A_7 \theta \nonumber \\
T_B&=&\imath A_8(\bar \epsilon \nu-\epsilon\bar\nu)\nonumber
\end{eqnarray}

\begin{eqnarray}
S_C &=&A_1+A_2\tau+A_3( \epsilon\theta+\bar\epsilon\bar \theta) \nonumber \\
V_C&=&\bar\epsilon A_4+\bar\epsilon A_5\tau+A_6\theta+\bar \epsilon A_7
\bar\theta \\
\bar V_C&=&\epsilon A_4+ \epsilon
A_5\tau+A_6\bar \theta+ \bar\epsilon A_7 \theta \nonumber \\
T_C&=&\imath
A_8(\epsilon \nu-\bar\epsilon\bar\nu)\nonumber
\end{eqnarray}

\begin{eqnarray}
S_D &=&0 \nonumber \\
V_D&=&\imath A_9\nu \\
\bar V_D&=&-\imath A_9 \bar\nu
\nonumber \\
T_D&=& A_{10}+A_{11}\tau
\nonumber
\end{eqnarray}

Here the parameters $A_1\div A_{11}$ are real numbers and the
kinematical variables  $\tau,nu,\bar \nu,\theta,\bar \theta$ are defined by
eqs.~(\ref{Tau}) above.

\subsubsection{The OPE graph contribution}

According to ref.\cite{19} the OPE graph contributes to $S_A,\> S_B, \>
S_C$ functions only, the corresponding expressions having the following
form.

\begin{eqnarray}
S_A&=&\frac{2g}{\tau-\mu^2}\bigl (G_0+G_1(\theta+\bar\theta)+
G_2(\theta\bar\theta)+G_3(\theta^2+{\bar\theta}^2) \bigr ) \nonumber \\
S_A&=&\frac{2g}{\tau-\mu^2}\bigl
(G_0+G_1(\bar\epsilon\theta+\epsilon\bar\theta)+
G_2(\theta\bar\theta)+G_3(\epsilon\theta^2+\bar\epsilon{\bar\theta}^2)
\bigr ) \\
S_A&=&\frac{2g}{\tau-\mu^2}\bigl
(G_0+G_1(\epsilon\theta+\bar\epsilon\bar\theta)+
G_2(\theta\bar\theta)+G_3(\bar\epsilon\theta^2+\epsilon{\bar\theta}^2)
\bigr ) \nonumber
\end{eqnarray}

Here $G_0,\ldots,G_3$ stand for free real parameters describing the
$\pi\pi$ interaction and the constant $g$ corresponds to the $\pi NN$
vertex.  It can be expressed via the familiar pseudovector coupling
constant $f$ ($f^2/4\pi\simeq 0.08$) as follows
$$
g=\frac{2m}{\mu}f
$$

\subsubsection{The $\Delta$ contribution}

In accordance with sec.2.1 the propagator of the $\Delta$--isobar
involves poles of the Breit--Wigner type:
\begin{equation}
\frac{1}{\omega_{\pi N}^2-M_{\Delta}^2+\imath M_\Delta
\Gamma_\Delta}.
\label{MMgl3}
\end{equation}

Here, $\omega_{\pi N}$ stands for the invariant mass of the $\pi
N$ pair. There are 6 such pairs:
\begin{eqnarray}
\omega_{pi}^2\qquad (i=1,2,3):\qquad (p+k_1)^2,\qquad
(p-k_2)^2, \qquad (p-k_3)^2, \\
\omega_{qi}^2\qquad (i=1,2,3):\qquad (q-k_1)^2,\qquad
(q+k_2)^2, \qquad (q+k_3)^2\,.
\label{MMgl4}
\end{eqnarray}

The numerators of the corresponding terms must have the correct
spin--isospin structure $(J_\Delta=I_\Delta=3/2)$; they have
to fulfill also the correct properties with respect to Bose and
crossing--symmetry requirements. The simplest way to coordinate
all these conditions
%among themselves consists of using the
is to use the
method of effective Lagrangians, the corresponding coupling
constants being regarded as free parameters.

Though we expect a pronounced energy dependence
($\Delta$--pole) only for the direct graph of fig.\,2d
(the additional nucleon -- and $\Delta$ --poles
appearing in graphs of figs.\,2a--2d cannot
give rise to the strong dependence of the kinematical
variables because they are located too far from the physical space)
we include in our analysis all graphs of figs.\,2a--2d (along
with crossing ones) for the sake of consistency.

To calculate the contributions of graphs
of figs.~2a--2d we use
 the following effective Lagrangians:

\begin{eqnarray}
{\cal L}_{\pi N N} &=& g_{\pi N N} \bar{N} \tau^a \gamma_\mu
\gamma_5 N \partial^\mu \pi^a ,
     \label{LpNN}\\
{\cal L}_{\pi N \Delta} &=&g_{\pi N\Delta}
\bar{\Delta}^a_\mu N \partial^\mu \pi^a + h.c.
 \label{LpND} \\
{\cal L}_{\pi \Delta \Delta} &=& g_{\pi\Delta\Delta}
\bar{\Delta}^a_\mu i \gamma_5 A^{abc} \Delta^{b\mu} \pi^c,
 \label{LpDD}  \\
{\cal L}_{\pi \pi \Delta N} & = & \bar{N}
( f_0 F_0^{abc} + 3 f_1 F_1^{abc}) i \gamma_5 \Delta_\mu^a
\partial^\mu \pi^b \pi^c + h.c.  \nonumber \\ & + &
\bar{N}( g_0 F_0^{abc} +
     3 g_1 F^{abc}_1)\gamma_\nu
\gamma_5 \Delta_\mu^a \partial^\mu \pi^b \partial_\nu \pi^c +
h.c.,\label{LppND}
\end{eqnarray}

where
\begin{eqnarray}
A^{abc}& = & 2 \delta^{bc} \tau^a
          + 2 \delta^{ac} \tau^b
          - 8 \delta^{ab} \tau^c
          + 5 \imath \epsilon^{abc},
          \nonumber\\
F_0^{abc}&=&i
 \epsilon^{abc} + \delta^{ab} \tau^c - \delta^{ac} \tau^b
\label{iso} \\
F_1^{abc}& =& \delta^{ab} \tau^c + \delta^{ac}
\tau^b - \frac{2}{3} \delta^{bc} \tau^a.\nonumber
 \end{eqnarray}
 are the invariant isotopic forms guaranteing the correct
 isospin (I=3/2) of the isobar field $\Delta^a$:
 \begin{equation}
 \tau^a\Delta_a=0 \label{tauD}
 \end{equation}
 ($\tau^a$ are the standard isospin matrices:
 $[\tau^a,\tau^b]=\imath\varepsilon^{abc}\tau^c$).

It should be stressed here that the choice  (\ref{LpDD})
of the $\pi\Delta\Delta$ coupling (non--derivative one) has
nothing to do with the problem of chiral symmetry breaking,
because (due to the equation of motion)
the chiral invariant form
of the $\pi \Delta\Delta$ coupling
$$
{\cal L}_{\pi\Delta\Delta}^{\prime} \sim
{\bar  \Delta}_\mu \gamma_{5}\gamma_\rho\Delta^\mu \partial^\rho\pi
$$
is equivalent
to that of the
eq. (\ref{LpDD}) on the mass shell of the isobar.
The off--shell
contributions renormalize the constants of the background and
the $\pi\pi N\Delta$ vertex. Since the parameters of the
background as well as those of the
$\Delta$-vertices
are considered to be free, the difference
 between "chiral" and "non--chiral"
types of  $\pi\Delta\Delta$  coupling becomes unimportant;
our choice of the $\pi\Delta\Delta$ vertex is dictated by simplicity.
The same is true for the   $\pi NN$ vertex also.
We use the
familiar pseudovector type (\ref{LpNN}).

This note also explains the absence
of the
terms with the derivative of the nucleon field
in  (\ref{LppND}) --- such a term
could again only renormalize the parameters already taken into
account.

The effective Lagrangians (\ref{LpNN}--\ref{LppND})
describe  the $\Delta$ contribution to the
$\pi N\rightarrow \pi\pi N$  amplitude (in the energy
region under consideration) in the most general way. They
contain 6 real free parameters ($g_{\pi
N\Delta},g_{\pi\Delta\Delta},f_0,f_1,g_0,g_1$), one of them
($g_{\pi N\Delta}$) can be fixed in accordance with the known
$\Delta\rightarrow \pi N$ decay width.
We leave it free due to
the reasons explained in sect.~2.1.

To complete the description of the formalism we give the form
of the $\Delta$ propagator used in our calculations:

\begin{eqnarray}
D_{\mu \nu}(p) &=& \frac{p_\mu \gamma^\mu
+M_{\Delta}}{p^2-M_{\Delta}^2+iM_{\Delta}\Gamma_{\Delta}}
S_{\mu\nu}, \label{Delp}\\
S_{\mu\nu}&=& -g_{\mu \nu} + 1/3 \gamma_\mu \gamma_\nu +
\frac{2}{3M_{\Delta}^2} p_\mu p_\nu
-\frac{1}{3M_{\Delta}}(p_\mu \gamma_\nu - \gamma_\mu p_\nu),
\label{Smn}
\end{eqnarray}
$M$ and $\Gamma$ being the isobar mass and
$\Delta \rightarrow \pi N$ decay widths respectively.

\subsection {The fitting procedure}

To fit experimental data  one has at first
to calculate the modulo squared
$\mid M \mid^2$
(averaged over initial and summed over final nucleon
polarizations) of the theoretical amplitude
for every channel of the reaction
(1) at any fixed values of the free parameters
$A_i$ ($i=1,...,21$).
The second
step consists of the integration of
$\mid M \mid^2$ over the appropriate part
of the phase space, the total energy $E$ being fixed.
This gives the theoretical expression:
\begin{equation}
\label{first}
\frac{\partial^{(r)} \sigma}
     {\partial x_1 ... \partial x_r}
\mid_{E=const.} = \sum_{m,n=1}^{21} A_m^* Q_{mn}^{(r)}
(E,x_1,...,x_r) A_n , \qquad \qquad (1\leq r \leq4).
\end{equation}
The distribution ${\cal D}$ to be fitted is given
by a convolution:
\begin{equation}
{\cal D}^{(r)}(E,x_1,...,x_r) \equiv
\frac{\partial^{(r)} \sigma}
     {\partial x_1 ... \partial x_r} (E,x_1,...,x_r)
\ast \Phi(E,x_1,...,x_r),
\label{MMgl5}
\end{equation}
where $\Phi$ is a known apparatus function
specifying the type
of distribution, experimental conditions, etc..
Expression (\ref{first}) is a bilinear form
in the free parameters $A_i$ with
the coefficient matrix $Q_{mn}^{(r)}$ being a function
of the set of variables
$(E,x_1,...,x_r)$
describing the distribution under consideration.

To perform the fitting procedure,
it is necessary to organize the
experimental distributions in the form
of a finite array of numbers.
To do this one has to divide the phase space  into a set
of N cells $C_i$ with "central points"
$P_i(E,x_1^i,...,x_r^i)$, with $i=1,...,N$.
Each point $P_i$
must be supplied with
the corresponding experimental number
$W_i$ (integral over the volume of $C_i$) and, of course,
with the suitable error bars.
In other words, one obtains a $r$--dimensional
histogram representing the experimental data in question.

Precisely the same procedure must be performed
for the theoretical distribution
on the right hand side of (\ref{first}). This means
that for each central
point $P_i$ of the distribution ${\cal D} ^{(r)}$ one must
prepare
the corresponding $21 \times 21$ matrix
$Q_{mn}^{(r,i)} (E,x_1^i,...,x_r^i)$, N matrices in total.
When several different experimental distributions are being
treated simultaneously, the necessary number of correlation
matrices increases proportionally.  Finally, to treat data at
different energies,
one has to prepare the above mentioned full
set of correlation matrices for each energy value.
Keeping
this in mind, one can easily imagine the size of computer
memory which is necessary to perform the data treatment.
(This
also explains why our program \cite{34} could not be run
earlier.)

After all preliminary steps have been done, the fitting
procedure is performed with the help of the standard program
FUMILI \cite{35}.

\subsection{Integration over the phase space }

The question of the integration over the phase space
is extremely
delicate. The crucial point is that  near the threshold
the phase space factor strongly depends on the values of
particle masses (this sensitivity was pointed out recently
by
R.E. Cutkosky --- see citation in ref.
\cite{21}).
The dependence
of the amplitude modulo squared is weaker
(corrections are
proportional to mass differences), but, nevertheless,
it must be taken into
account when one uses the standard
version of the Monte Carlo program FOWL
\cite{36} (or any other routine)
for calculation of the integrals.

Here, we specify the remark by R.E. Cutcosky
and show that the
amplitude modulo squared contains
two different parts, one of which requires
special caution with respect to the values of masses
used in the
course of  computation.
To explain this statement we
must return back to expressions (\ref{X1}) and (\ref{X2}).
According to these formulae, the amplitude modulo squared can
be written in the form:
\begin{equation} \label{quad}
\frac{1}{2} \sum_{\mu,\nu=1}^{2} \mid X^{\mu\nu} \mid^2 =
\sum_{i,j=1}^4 X_i^* M_{ij} X_j , \end{equation} where $ X
\equiv (S,\bar{V},V,T)$ and the matrix $M$ is hermitian:
\begin{eqnarray}
\label{Mlm}
M_{lm} &=& \frac{1}{2} Tr \{ \tilde{\Gamma}_l
(\hat{q}+m_f) \Gamma_m (\hat{p}-m_i)\};
\nonumber \\
\Gamma & \equiv& \{I,\hat{k},\hat{\bar{k}},\frac{i}{2}
[\hat{k},\hat{\bar{k}} ] \}, \qquad
\tilde{\Gamma} \equiv \{I,\hat{\bar{k}},\hat{k},
\frac{i}{2}
 [\hat{\bar{k}},\hat{k} ] \},
\end{eqnarray}
with $\hat{r}=\gamma_{\mu}r^{\mu}$ for $r=p,q,k,\bar{k}$;
$m_i$ and $m_f$ are the masses of incoming
and outgoing nucleons, respectively.

The matrix $M$ degenerates at the boundary
of the phase space and
the quadratic form (\ref{quad})
loses the property of positive
definiteness outside of the phase space.
The bounds of the phase
space domains determined at the physical
values of masses
$m_i\neq m_f$
and at the average ones
$m_i=m_f=\bar m$ do not coincide.
This
means that one must use the physical masses
along with momenta
%generated by the FOWL program
to compute
both the phase space
factor and the matrix elements $M_{ij}$.
 Otherwise, incorrect and even
negative contributions to the left hand side of (\ref{quad}) are
unavoidable. The point is that errors
must be compared not with the particle masses,
but with the
distances from the boundaries of the real physical phase
 space, the latter being small if the energy is low.

It is  easy to understand
that all above arguments are valid
for the treating the mass differences of the
pions also.

Thus, we conclude that
one must use the true mass values both for
the phase space factor and for the matrix elements
(\ref{Mlm}) to avoid problems with the kinematics near
threshold.  One can take the averaged values
$\bar{m}$ for the nucleon and $\bar{\mu}$ for the pion masses
for the computation of the amplitude vector $X$ in (\ref{quad});
in this case the errors will be
of the order of the isospin symmetry breaking.

\subsection{Description of the database}

In our analysis we use the full set of published (up
to May 1994) experimental results on total cross sections of
5 channels (namely,
$\pi^-\pi^+ n$,
$\pi^-\pi^0 p$,
$\pi^0\pi^0 n$,
$\pi^+\pi^0 p$,
$\pi^+\pi^+ n$)
of the
$\pi N \rightarrow \pi \pi N$ reaction
at $P_{lab}\leq 500 MeV/c$.
These data (105 points in total) along with the corresponding
references are listed in the Table 1. For convenience
%the sake of visuality
the same  data are presented graphically on Figs.3a~--~3e in
the form of the r--th (r=1,...,5) channel quasi--amplitude
$\tilde{M_r}$ dependence on the incident beam momentum~$p_{lab}$.
The quasi--amplitude is defined as follows
\begin{equation}
\tilde M_i=\{\frac{\int \frac12 \sum_{\mu\nu=1}^2 |M_i^{\mu\nu}|^2
d\Gamma}{\int d\Gamma }\}^{1/2} \sim \{\frac{\sigma_{tot}}{\int
d\Gamma}\}^{1/2}
\label{MMgl10}
\end{equation}
(see Table~1),
where $d\Gamma_3$ is the standard element of the 3--particle
phase space volume. This method is much more convenient than
the one conventionally
used ($\sigma_{tot}$ as a function of
$p_{lab}$), because the extremely rapid
energy dependence of the numerator caused by the phase space
volume is cancelled by the denominator.
The resulting smooth energy dependence of the quasi amplitude
$\tilde M_i$ provides a fairely direct insight in the
quality of the various data points. As an immediate consequence
the point no. 12, being obviously inconsistent, was excluded
from the data set analysis from the very beginning.

\subsection{Strategy of the fitting}

To answer the questions raised in sec. 1,
we choose the following procedure:
first, we  fit  the total cross
section data in order to see what we can learn
from this information alone. To test the
sensitivity of the data to the OPE
or $\Delta$ contribution, we begin to fit with
the background amplitude B only
(11 parameters). For comparison with models it is interesting
also to fit these data using the OPE amplitude only (4
parameters) and, for completeness and  to test the
sensitivity, using the $\Delta$ amplitude only.  At this point
it has to be stressed, that the amplitude $\Delta$ alone is not
the full amplitude which can be calculated from the diagrams,
since its  smooth part corresponding to the off--shell
$\Delta$ interaction is included into B.  In contrast, the
amplitude $\Pi$ itself has a clear physical meaning because
it  corresponds to the simplest variant of the OPE model (see
ref.
\cite{19}).

More physical, however, are the results
 of a fit with a combination
of two of these amplitudes, like $B+\Delta$ or $B+\Pi$.
At the end, the fit including the full
 amplitude with 21 parameters is performed.
The full amplitude, according to sec. 2.1,
should be general enough to describe
every reasonable collection of $N$ data points
 in the momentum region in question.
Any set of data points tolerating
  a reasonable fit can therefore
be considered as being selfconsistent.

The quality of a fit is considered to be  good if the value
${\bar\chi}^2 \approx 1$, where
\begin{equation} \label{chisquare}
{\bar\chi}^2\equiv\frac{1}{M}\sum^N_{i=1}\chi_i^2,
\end{equation}
with $N$ being the numer of datapoints, whereas $M$
being the number of degrees of freedom ($N$ minus
number of fit paramaters).
 When there are some data  points
  with individual contributions of $\chi^2_i$ more than 5
(which is of course our
own, arbitrary upper limit), these points
need a closer look and have been
excluded for the next step of the fit.

After the exclusion of some data points
the fit with the full amplitude
and some fits (leading to a large ${\bar\chi}^2$ before)
for a reduced set of parameters
are repeated to see the influence
of these data points on the results.

\section {Analysis of data on total cross sections}

As mentioned above our analysis of the
$\pi N\rightarrow\pi \pi N$ reactions is based mainly on the results from
ref.~\cite{19}, supplemented by the inclusion of the $\Delta$-isobar.
Even in this fairly restricted parametrization the model -- as
reflected by its dependence on 21 parameters -- is highly complex; a finding,
which is stressed by the result of the fitting procedure from section 4,
that none of the various ingredients from eq. (9) of the model
strongly dominates the complex dynamics  at low energies.
Due to this subtle interplay we expect that the various
$\pi N\rightarrow\pi \pi N$ isospin channels exhibit quite different
features, such as in their extrapolation to the corresponding
threshold or in their angular distributions.
To gain deeper insight into our subsequent results and to facilitate their
interpretation, we study in appendix A ``quasi'' $\pi N$ elastic scattering
(i.e. equal nucleon and pion mass)
as a much simpler system with related dynamics.
This consideration will help
to understand better the logic of the subsequent analysis of
the
$\pi N\rightarrow\pi\pi N$
reaction.

Our analysis of the $\pi N \rightarrow \pi N$ reaction at low energies
exhibits
the following conclusions:
\begin{enumerate}
\item
Dealing with reactions of particles with spin
one has to take account of the existence of several independent
scalar amplitudes. In such a case the term"simplest dynamics"
loses its transparency and must be specially defined.
\item
The spin structure of the amplitude manifests itself even in
the case
%when one deals with
of unpolarized particles: it causes the
"minimal" (unremovable) dependence of the distributions on
kinematical variables (due to the presence of matrix $G$ ---
see eq.~(\ref{i27})). It can be shown that it is impossible
to construct a physically correct amplitude which would
result in the precise "phase space-like" behavior of
differential cross-sections in all channels (\ref{i26}).
\item
The explicit form of the "minimal" kinematical
dependence mentioned above is determined by the interplay of the values of
parameters describing the independent scalar amplitudes $A_i$.
It may be rather complicated even in the "simplest" cases when
all $A_i$ are constants. In what follows we call this phenomenon
the "spin correlation".
\item
The isospin conservation law connects, to some extent, the spin
correlations in different channels. Then, crossing
symmetry and Bose statistics give the additional limitations on
the admissible form of the latter correlations. Altogether,
these symmetries result in the existence of spin--isospin --
Bose -- crossing correlations connecting the kinematical
dependence of the amplitudes of different channels.
It is easy to understand that the more identical
particles are involved in the reaction the stronger are these
correlations.
\item
The low energy behavior (the form of the extrapolation law) of
the total cross section in a given channel depends on the relative
magnitudes and signs of the amplitude parameters; it may be
different in different channels (see Appendix, Ex.III). To get
the form of the extrapolation law one has to analyze the data
on distributions at several values of incident momentum (see
Appendix, Ex.I--III).  Example IV shows that the "natural" terms (that of
order $k^0$) may be absent in low energy expansions of total cross sections
of all channels.  \end{enumerate}

Keeping in  mind these basic findings from the
the ``quasi'' $\pi N$ elastic scattering we can
start now with the analysis of the modern database on total
cross sections of the more complicated
$\pi N\rightarrow\pi\pi N$                  reaction
in the low energy domain $P_\pi\leq 500 MeV/c$

\subsection{General theoretical consideration.}

First of  all, let us summarize some formulae concerning the
$\pi N\rightarrow \pi\pi N$
reaction (hereafter we use the notations of ref.~\cite{19}).
The five experimentally accessible
channels of this reaction along with the corresponding
amplitudes $X_r$ and statistical factors $f_r$ are listed
in Table~2 (in comparison with ref.~\cite{19} the numbers of
two last channels are changed).

In accordance with the eq.~(\ref{Mlm}) the total cross section
of the $r$-th channel can be written in the form
\begin{equation}
\sigma_r=N\int d\omega\cdot f_r\cdot{X_r}^\dagger M X_r
\label{g32}
\end{equation}
 Here, $N$ is the normalization factor (see eq.(57) of
 ref.~\cite{19}), $X_r$ --- the amplitude vector of $r$-th
 channel, i.e.
\begin{equation}
 X_r\equiv (S_r,\bar V_r, V_r, T_r)  \label{g33}
\end{equation}
 and the hermitian $4\times4$ matrix $M$ (which is the same for all
 channels, if one neglects the isospin symmetry breaking) has the
 elements given by eq.~(\ref{Mlm}).

 With the help of eq.~(\ref{g32}) it is easy to analyze the
 correspondence of some statements concerning the dynamics of
 the reaction in question with the experimental data on total
 cross sections.

 I. The OPE -- dominance hypothesis.

 According to popular wisdom the main contribution
to the low energy
 $\pi N\rightarrow\pi\pi N$
amplitude originates from the OPE--graph ($\Pi$; fig.1). It is
easy to check that in this case (we assume the leading order of
$4\pi$--interaction and the arbitrary form of the nucleon form
factor) the amplitude of the reaction (\ref{piN}) takes the
form:
\begin{eqnarray}
S_A &=& S_B=S_C=S(\tau), \qquad S_D=0  \nonumber \\
V_X &=& \bar V_X=T_X =0, \qquad (X=A,B,C,D; eq.(\ref{Amp}))
\label{g34}
\end{eqnarray}
Here, $S(\tau)$ depends solely on $\tau$ (see eq. \ref{Tau}). Using the
expressions from Table~2 and eq.(\ref{g32}), one obtains
the following relations among the total cross sections of five
basic channels:
\begin{equation}
\frac18\sigma(\pi^{-}\pi^{+}n)=\sigma(\pi^{-}\pi^{0}p)=
\sigma(\pi^{0}\pi^{0}n)=\sigma(\pi^{+}\pi^{0}p)=\frac14\sigma(\pi^{+}\pi^{+}n)
\label{g35}
\end{equation}
The quality of this relation is expected to improve with the dominance
of the OPE graph.
%These relations are expected to be the more precise, the less
%significant is the total contribution of the other (than OPE)
%graphs.
Of course, in the vicinity of threshold
($p_\pi\leq 300 MeV/c$)
one must take into account the isospin breaking (see Sec.2.4
above); in this case it would be more correct to consider the
corresponding relations for  quasi-amplitudes.

Even a short glance at the Table~1 is enough to realize that
the relations (\ref{g35}) strongly contradict the experimental
data in the area of
$p_\pi\leq 500 MeV/c$.
As a consequence the OPE--dominance hypothesis is questionable and the
 straightforward application of the Chew-Low extrapolation
 procedure cannot result in reliable values of $\pi\pi$
 parameters. This statement is now commonly agreed upon.
\vspace{0.5cm}

II. The $S$--wave final state.

In accordance with the modern data on distributions
[7--9,17,25] at
$p_\pi\leq 380 MeV/c$
each pair of final particles in reaction (\ref{piN}) is
expected to be in a
relative $S$--wave state. As it was shown in \cite{67}, in
this case some of relations (\ref{g35}), namely
\begin{equation}
\sigma(\pi^{-}\pi^{0}p)=\sigma(\pi^{+}\pi^{0}p)=\frac14\sigma(\pi^{+}\pi^{+}n)
\label{g36}
\end{equation}
are expected to be valid. The comparison of (\ref{g36}) with
the corresponding data from Table~1 shows, however, the
striking disagreement. Since both parts of the picture
(distributions and total cross sections)
contain the same dynamics
%are connected solely with the experimental data
we are forced to conclude that there
is either an implicit
 contradiction between the data on distributions and those
on total cross sections or that the assumption on $S$--wave final states
is too restrictive.

It should be noted here that one of the relations
(\ref{g35},\ref{g36}):
\begin{equation}
\sigma(\pi^- \pi^0 p)=\sigma(\pi^+ \pi^0 p)
\nonumber
\end{equation}
seems to be supported by data. This relation implies (see
Table~2) the smallness of the isotopic amplitude $D$ in
(\ref{Amp}) in comparison with $C$. If this is true,
all  distributions in both channels must be
approximately equal. From the purely theoretical point of view
the smallness of $D$ would  imply a delicate cancellation
of various contributions. That is why it would be extremely
interesting to check this point at momenta
$p_\pi\geq 400 MeV/c$
where the existing data are still scarce.
\vspace{0.5cm}

III. The phase--space like distributions.

Before starting the discussion we would
like to stress that our consideration here is only preliminary;
as already stated in our introduction,
the problems concerning the distributions will be discussed in
detail in forthcoming publications.

In accordance with the results of ref.~\cite{6,7,8,9,17,25} none of the
distributions in
four channels (except $\pi^-\pi^+ n$) up to
$p_{\pi} < 320MeV/c$ show (within the error
bars) any deviation from the purely
3-particle phase space kinematics.  What does this fact tell us?

To answer  this question it is necessary to use
eq.~(\ref{i31}) along with the explicit expressions for the
matrix elements $M_{ij}$ ($i,j=1,..,4$).
A short look at eqs.(56) of ref.~\cite{19} shows that
almost \underline{each} element of the matrix $M$  depends,
for example, on
$\tau$. Thus to remove this ("necessary") $\tau$--dependence
one has to select the suitable amplitude vectors $X_r$.
 The direct calculation, however, shows that this is
impossible: no physically reasonable choice of $X_r$ can
ensure the exact independence of $|M_r|^2$
of all four kinematical variables. This result
becomes more clear from the examples considered in
Appendix and the explicit form of the
OPE--contribution.

 Thus, we are forced to conclude that the experimental
observation of the "phase--space like" behavior of
distributions  would show that the precision
of the data is insufficient to display the
$\tau$ -- dependence originating from the OPE -- graph.
 In fact it shows that the experiment cannot
distinguish between spin 1/2 and spinless nucleons.
Such
precision, of course, does not provide the opportunity to
extract reliable information on the details of the dynamics
of the process under consideration. It can also result in
large uncontrollable errors if the total cross section
is obtained with the help of integration of distributions over
the corresponding variables.

So, even a very general consideration shows that the data
set on total cross sections may contain
insufficient information for the extraction of some
parameters of the amplitude. Moreover, some of the data points
(especially, those obtained with the help of the integration
procedure) may be incorrect in addition
to
other -- purely experimental -- reasons.

\subsection{Results of the Fit}

In accordance with the general strategy of
the data fitting as discussed in Sec.~2.6
we have performed 7 different fits, each one
corresponding to the specific set of the amplitude
parameters involved:
$B,\ \pi,\ \Delta,\ B+\pi,\ B+\Delta,\ \pi+\Delta,\ B+\pi+\Delta$.
Some results of the best fits are given in Table~3. It should be
noted here that for each kind of fit there were found several
solutions with approximately the same values of $\chi^2$; they
differ from each other by the parameters values. Thus, in
Table~3 we give the results corresponding to only one "typical"
solution for each kind of fit.

%The Table~3 is organized as follows:\\
%Column 1:\ \ \ the kind of the fit.\\
%Column 2:\ \ \ the total value of $\chi^2$ corresponding to
%one  of the best solutions.\\
%Column 3:\ \ \  the value of $\bar\chi^2$\\
%Column 4:\ \ \  the list of the "doubtful points" (DP's);  numbers
%are given in accordance with Table~1. The point $P_q$ is called
%"doubtful" if $\chi^2_q\geq 5$.\\
%Column 5:\ \ \  the summary contribution $\chi^2(DP)$ of all DP's
%listed in column~4 to the total value of $\chi^2$.

The results of the primary ``filtration'' of the data sample (exclusion of
``doubtful data points'', DP)
show that the latter contains some DP's. Indeed, 7 points ---
namely, 6, 64, 73, 75, 88, 91, 93 --- appear in \underline{all}
solutions as doubtful ones. The point $no.$12 omitted from the
very beginning must be also added to this list. These 8 points
need special verification. Until the latter is not done, they
must be excluded from the list of the data used for the
subsequent analysis. Excluding them --- the database is reduced
by this procedure from 105 to 97 points --- we can perform the
second step of the data filtration to elucidate the status of 4
points ($nn.$ 1, 4, 76, 90) which appear as doubtful in some
solutions. The results of this --- second ---step are given in
the Table~4. It is organized in the same manner as the
Table~3; the points numbers placed in brackets indicate that the
corresponding point gives the individual contribution
$\chi^2_i$ close to the critical value ($\chi^2_i=5$)\\
%(We watch specially the points $nn.$ 1, 4, 76, 90).

The consideration of Table~4 shows that two points -- $nn.$ 1,4
-- which appeared in Table~3 as "almost doubtful" can be
regarded as  being really doubtful.
%It can be added that there
%was not found any solution (with reasonable $\chi^2$)
For them we found no solution with
$\chi^2_i\leq 4, \ \  (i=1,4)$.

Summarizing, the list of doubtful data points contains:
\begin{eqnarray}
\label{r37}
& &\pi^+\pi^- n\ \ \  \mbox{\rm channel} \hspace{3cm} 1, 4, 6, 12 ;
\nonumber\\
& &\pi^0\pi^0 n\ \ \  \mbox{\rm channel} \hspace{3.2cm} 64, 73, 75; \\
& &\pi^+\pi^+ n\ \ \  \mbox{\rm channel} \hspace{3cm} 88, 91, 93,
\nonumber
\end{eqnarray}

It should be noted that all DP's are concentrated in channels
with the neutron in the final state; the two remaining channels
(those with the final state proton) do not contain DP's.

The database filtration procedure could be continued further,
but we prefer to stop it here, because in the next paper we
shall include the additional experimental information (the
data of \cite{58}
on distributions in $\pi^-\pi^+ n$ -- channel) to get
more reliable and well grounded results. It is necessary to
note here that we do not regard the list (\ref{r37}) as the
final one. Unfortunately the "TRIUMF--OMICRON problem" \footnote
{
The origin of the "TRIUMF--OMICRON problem" has been recently pointed out by
Ortner et al. \protect \cite{11}. For the OMICRON collaboration the total cross
sections have been obtained from angular distribution data in a
restricted angular range assuming a smooth continuation over the full phase
space.
Angular distributions for the $\pi^- p \longrightarrow \pi^+ \pi^- n$ channel
\cite{14} do not support such a simple extrapolation.
}
in the
$\pi^+\pi^+ n$--channel
(which is clearly visible on fig.3e)
 is not as trivial as it seems at the
first glance. The attentive reader could notice that the point
$no.$90 (TRIUMF) oftenly appears as the doubtful one along  with
the points $nn.$ 88,91,93 (OMICRON) -- at least in the best
solutions. Our experience prompts that it is the hint on the
impossibility to resolve the above mentioned problem with the
help of purely statistical methods on
the database in question: it is necessary to use
additional independent information. As to our list of DP's in
$\pi^-\pi^+ n$-- and   $\pi^0\pi^0 n$--channels, we believe it
is correct and the new information can only display some new
DP's in addition to those quoted  above.

Our conclusion about the presence of DP's in the data sample in
question contradicts the result derived by Burkhardt and
Lowe (BL) in refs.~\cite{18,68} where the authors have stated
that the database on total cross sections is internally
consistent up to
$p_{lab}=400MeV/c$ (apart from a momentum region $310\,-\,370MeV/c$
in the $\pi^{+}\pi^{+}n$--channel).
It is interesting to elaborate on this aspect in more detail.

In what follows we discuss only the latest results published in
ref.~\cite{68}, where the new data \cite{25} on total cross sections
of the $\pi^+\pi^0 p$--channel are taken into account.
In accordance with the  prescription of ref.~\cite{68} all data
points in $\pi^+\pi^+ n$--channel corresponding to the momentum
interval
$p_{lab}=310\,-\,370 MeV/c$
(totally 7 points) must be excluded from the fitting procedure.
Then, in their work Burkhardt and Lowe use the reduced data
set: points $nn.$ 4, 6,  7, 9, 11, 12,  16, 17, 41,  64, 70, 72,
74 are not included in their database. Thus, to compare the
results, we have also excluded the points mentioned above and
performed the fitting of the remaining database (49 points).
Then, using the parameters corresponding to the best fit (or,
if several, to one of them) we have computed the individual
contributions $\chi^2_i$ for all points in the data set,
including those previously omitted. The values of $\chi^2_i$ in
the BL fit,
which are not listed in ref.~\cite{68},
  have  been obtained with the help of the interpolation
of the numbers given in the Table~1 of ref.~\cite{68}. The
corresponding results concerning the DP's (\ref{r37}) are given
in Table~5, where we use the parameters of one of the best
solutions with the background amplitude $B$ only (point $no.$12, as
noted above, is omitted from the very beginning).

The comparison of numbers in Table~5 shows that the points in
question look inconsistent in both fits. This result confirms
our list  (\ref{r37}) of DP's.

It must be noted also that in refs.~\cite{18,68}
points nn.\,73, 75 have been omitted along
with those quoted above, while near threshold points nn.\,1, 3, 5
(and some others corresponding to higher momenta) were supplied with
the beam momentum values considerabley different
from those quoted in the original papers, this latter difference
originating from the additional corrections of purely experimental
character which were taken into account in refs.~\cite{18,68} and
neglected in our analysis because we use only published
results. (We are grateful to Prof.~J.Lowe for a detailed
discussion of this question). In addition, the systematic errors in the
$\pi^{0}\pi^{0}n$--channel were handled in a different way in
refs.~\cite{18,68}. The differences pointed out above between the two data
samples give rise to the different corresponding conclusions.

It should be mentioned that if the corrected values of effective beam momentum
(in accordance
with BL data base) are used in our
analysis, only the point no. 1 (corresponding to $p_{\pi}=302.5\,MeV/c$
in the BL sample) looks considerably better and must be deleted from the
list (\ref{r37}). All other results remain practically unchanged.

{}From our findings above a detailed discussion of the parameter
dependence of the
model would be premature.
This can be best illustrated by the mutual comparison of numbers from
Table 6 where we give the  typical examples of solutions obtained in
a course of the pruned database fitting with different groups of
parameters. The considerations of this table shows that every statement
on the values of $\pi\pi$--parameters based on the analysis of the data on
total cross sections only would be strongly model dependent because the
present knowledge of the background processes (and, hence, the values
of $A_1,\ldots,A_{11}$) is too pure, the latter processes playing a very
important role in the $\pi N\rightarrow\pi\pi N$ reaction mechanism.

\section{Conclusion}

\indent
Before the discussion of our results we would like to start with some
notes concerning the adequate treatment of the experimental
data on
$\pi N\rightarrow\pi\pi N$
             reaction in the momentum region in question
($p_{\pi}\leq 500 MeV/c$).

Even a short glance at Figs.~3a--3e is enough to recognize
that the modern database contains some pairs of mutually
inconsistent points. Thus, the preliminary selection (or, the
same, filtration) of data is unavoidable if one intends to
extract  with the reasonable accuracy such quantities as
$\pi\pi$--scattering parameters. The only question is how to
do this selection.

Of course, one can simply exclude the mutually contradictory
pairs of points from the database as it was done in
refs.~\cite{18,68}. Unfortunately, such a recipe does not
guarantee the consistency of the remaining collection of
data. Indeed, after removing "doubtful pairs" one
still has 5 independent subsets of data points
(the total cross sections of 5
channels).
These are related by isospin and Bose symmetries, and it is
essential that these symmetries are incorporated in
the analysis. The channels are also related by crossing symmetry
(which was ignored in the analysis of refs.~\cite{18,68}).

Thus, it is evident that any correct data selection
must take into account all symmetry requirements. If this
condition is fulfilled one can use a wide range of equivalent
approximants for the ``data filtration''. Their specific choice
(polynomials, spherical harmonics, etc.) can have influence only
on the number of free parameters necessary to achieve a good description
of the data. From this point of view our choice
($B+\Pi+\Delta$) for the
$\pi N\rightarrow\pi\pi N$
transition amplitude
seems to be the most economic one. In addition
to the symmetry requirements it takes account of rather general
features of the   process in question: the existence of a
smooth background $B$ and the $\pi$ and $\Delta$ poles close to
the kinematical regime investigated.

Of course, we could use polynomials of higher degree for the
background contribution (this would introduce $\sim 30$ new
parameters for the second degree polynomials). Our results
show, however, that this is not necessary, because a
reasonably good description of data is achieved with
linear approximants (in our next paper it will be shown that
this is true also for the data on distributions).

Summarizing we conclude that the procedure used here for the
primary filtration of data is general enough: in the case when
some data points cannot be described with the reasonable
accuracy, they need further experimental verification. Until
such a verification is not done these points must be excluded
from the database.

With this in mind this conclusion we can make now
the following statements.
\begin{itemize}
\item
The modern collection of the experimental data on total cross
sections of 5 channels of
$\pi N\rightarrow\pi\pi N$
reaction at $p_{\pi}\leq 500 MeV/c$ contains some doubtful
points (as listed in eq. (\ref{r37})) which need special verification.
%The corresponding
%list includes (among others) the following 10 points (the
%number are given in  accordance with Table~1):\\
%$\pi^+\pi^- n$ channel:\hspace{2cm} 1,4,6,12 \\
%$\pi^0\pi^0 n$ channel:\hspace{2cm} 64,73,75 \\
%$\pi^+\pi^+ n$ channel:\hspace{2cm} 88,91,93
\item
The experimental data on two channels with a final state
proton ($\pi^-\pi^0 p, \pi^+\pi^0 p$) are in statistical sense
reliable: they do not  contain DP's. Nevertheless, it would be
extremely useful to perform new accurate measurements of the
$ \pi^+\pi^0 p$ channel to increase its statistical
significance. It would be of special interest to perform such
measurements at relatively large values of the incident
momentum: $p_{\pi} = 400\,-\,500 MeV/c$. The corresponding
results would give the opportunity to check the approximate
equality of cross sections of these two channels.
\item
The pruned database (i.e. that obtained after the exclusion of
DP's) can be described in terms of our ansatz with a reasonable
value of $\bar\chi^2$ ($\leq 1.2$) (see Fig. 3).
Unfortunately, there exist
at least several quantitatively equivalent solutions
corresponding to rather different values of parameters (the
correlations among them are extremely strong). Thus, the experimental data in
question do not allow the extraction of a
%provide the opportunity to extract the
unique
set of parameters in the $\pi N\rightarrow\pi\pi N$.
%It is possible to make
Only
qualitative conclusions concerning the influence of the various
reaction mechanisms are possible from this limited set of data.
\item
The most important contribution comes from the background
term $B$. This term alone admits solution with
$\bar\chi^2=1.4$
The addition of the  $\Pi$ term improves the
description significantly:
$\bar\chi^2(B+\Pi)=1.2$.
This fact shows that --- in accordance with  common
expectations --- the OPE term plays an essential role, its
contribution is definitely visible even on the database in
question and, therefore, the corresponding parameters might be
extracted if the database is enlarged by the inclusion of the
accurate enough data on distributions. The role of $\Pi$ term,
however, is by no means dominant
($\bar\chi^2(\Pi)=2.3$),
so the well known Chew-Low extrapolation procedure seems
questionable. This finding coincides well with the  results
of refs.~\cite{61,62,60,59}.
 \item
 In contrast to the $\Pi$ term, the $\Delta$
 contribution is rather insensitive on the database in question.
 The only observation which is of some interest can be formulated as
 follows: the most significant $\Delta$ contibution corresponds to
 the parameters $g_0$ and $g_3$ appearing in graphs of Fig.~2d. No definite
 statements concerning the other four parameters associated with the
 mechanism of the isobar exchanges (graphs of Figs.~2a--2d) can be done.
 In particular, the values of $g_{\pi N\Delta}$ and $g_{\pi \Delta\Delta}$
 cannot be fixed from the data on total cross sections.

 To  study the $\Delta$ mechanism one needs the more extensive
 database which must contain the data on distributions at
 several beam momenta.
 \item
 The informative capacity of the modern database on total cross
 sections is absolutely insufficient for the model independent
 extraction of the parameters describing the low energy $\pi\pi$
 interaction. The widely used statement that these parameters
 are most sensitive on and, thus, best extracted from the data on
 total cross sections near threshold, which is based solely on
 the Olsson--Turner model, is misleading.
 For the extraction of reliable information on $\pi\pi$
 interaction it seems much more reasonable to use the accurate
 data on distributions at several values of incident momentum
 in the interval $300\,-\,500 MeV/c$ (see ref.~\cite{19}).
\end{itemize}

Summarizing the present status of our analysis we conclude that
information from the total $\pi N \, \rightarrow \, \pi \pi N$
cross sections only rather quantitatively fixes the basic features of the
dynamics. Evidently more sensitive data, such as
angular distributions, are crucial for a
quantitative understanding.
The
analysis of the enlarged database containing
data both on total cross sections of 5 channels and on
distributions in the $\pi^+\pi^- n$ channel
is presently still troubled with serious numerical problems.
It will be discussed in a forthcoming paper.

\section{Acknowledgements}

We are greatful to A.A.~Bolokhov for the critical
reading of Sec.2 and the discussion of questions concerning
the $\Delta$ isobar contribution, G.A.~Feofilov, G.A.~Leksin,
J.~Lowe,
D.~Malz, R~.M\"uller, H.--W.~Ortner, O.O.~Patarakin, M.V.~Polyakov
and A.I.~Schetkovsky for the numerous valuable discussions
of general aspects of the subject and D.Pocanic for sending us
the information on experimental results prior to publication.
Two of us (S.G.S and V.V.V) are glad to use the opportunity to thank
all our German  colleagues for their friendly support and warm
hospitality during our visit at Erlangen University where the most
part of this work was done.

\newpage

\appendix

\section{Appendix: ``Quasi'' $\pi N$ elastic scattering}
\setcounter{equation}{0}
\renewcommand{\theequation}{A.\arabic{equation}}

As an illustrative example on the subtle influence of different dynamical
elements on the extrapolation and the angular distributions
in different isospin channels,
let us consider the familiar process: ``quasi'' $\pi N$ elastic
scattering:
\begin{equation}
\pi^a (k)+N_{e}^{\nu}(p) \rightarrow \pi^{b}
 (k')+N_{f}^{\mu}(p')  \label{i20}
 \end{equation}
 The term "quasi" used above means that --- to simplify
 the reaction kinematics --- we consider the hypothetical case of
 equal pion and nucleon masses
$ k^2=p^2={k'}^2={p'}^2=m^2 $
In (\ref{i20}) $a,b=1,2,3$ and $e,f=1,2$ are the isotopic
indices, $\mu,\nu=1,2$ the nucleon polarizations.

The amplitude of the process (\ref{i20}) can be represented in
the following form (for a comprehensive review see
ref.~\cite{66}):
\begin{equation}
 T_{ba,fe}^{\mu\nu}=
\delta_{ba} \delta_{fe}\  T^{\mu\nu (+)}+
{\imath \varepsilon}_{bac}
(\tau_c)_{fe}\ T^{\mu\nu (-)}
\label{i21}
 \end{equation}
Each of the isotopic amplitudes  $T^{\mu\nu (\pm)}$ can be
written as follows:
\begin{equation}
  T^{\mu\nu (\pm)}={\bar{u}}^{\mu}(p') \bigl \{
 m A^{\mu\nu (\pm)}+ \hat Q  B^{\mu\nu (\pm)} \bigr \}
 {\bar{u}}^{\nu}(p)
 \label{i22}
 \end{equation}
here $\hat Q =(\hat k + \hat k')/2$,
$A^\pm$ and $B^\pm$ being the functions of scalar kinematical
variables:
$t$~and~$\nu=(s-u)/4$.
The coefficient $m$ at $A^\pm$ in (\ref{i22}) is introduced for
dimensional reasons.
% the equalizing of $A$-- and $B$-- dimensions.
The crossing and Bose symmetries imply:
\begin{eqnarray}
A^\pm(\nu,t) &=& \pm A^\pm(-\nu,t) \nonumber \\
B^\pm(\nu,t) &=& \mp B^\pm(-\nu,t) \label{i25}
\end{eqnarray}

Formulae (\ref{i21}--\ref{i25}) determine the so-called
"canonical" form of the amplitude
 $T_{ba,fe}^{\mu\nu}$ of the process (\ref{i20}).

 Let us consider now three experimentally accessible channels of
 the reaction (\ref{i20}) and write down the corresponding
 amplitudes ($T\equiv A,B$):
\begin{eqnarray}
1. \ \ \pi^+ p\rightarrow \pi^+ p &\hspace{2cm}&
T_1=T^+ -\frac12T^-  \nonumber \\
2. \ \ \pi^- p\rightarrow \pi^- p &\hspace{2cm}&
T_2=T^+ +\frac12T^-  \label{i26} \\
3. \ \ \pi^- p\rightarrow \pi^0 n &\hspace{2cm}&
T_3=-\frac{\sqrt{2}}{2}T^-  \nonumber
\end{eqnarray}
Evidently, the amplitude $T_r$ of the $r$-th channel can be written in
the form of eq. (\ref{i22}) with $A_r, B_r$ being constructed in
accordance with eqs.(\ref{i26}).

Suppose, further, that we analyze the data on three channels
(\ref{i26}) with unpolarized nucleons. In this  case the
amplitude modulo squared (summed over the final and averaged
over the initial nucleon polarizations) has the form:
\begin{equation}
|T_r|^2=\frac12 Sp\bigl \{
(\hat p +m)(mA^\dagger+\hat Q B^\dagger)
(\hat p' +m)(mA+\hat Q B) \bigr \} \nonumber
 \end{equation}
 or, equivalently:
\begin{equation}
|T_r|^2= \sum_{k,p=1,2} {V^r_k}^\dagger G_{kp}{V^r_k}
\label{i27}
 \end{equation}
here, ${V^r}\equiv ({A^r},{B^r})$ is the "amplitude vector" of
the $r$-th channel and the real and symmetric matrix G has the
following elements:
\begin{equation}
G_{11}=m^2(4m^2-t),\
G_{12}=G_{21}=4m^2\nu, \
G_{22}=4\nu^2+\frac14 t(4m^2-t)
\label{i31}
\end{equation}

Thus, the differential cross sections for the $r$-th channel
may be written as follows:
\begin{equation}
d\sigma_r=N\cdot {V^r}^\dagger G{V^r}\cdot dt
\label{i29}
 \end{equation}
where $N$ contains all "trivial" kinematical and normalization
factors.

%Even without a partial wave expansions for the
%amplitudes $A^\pm$ and $B^\pm$
Let us suppose that
it is possible to make use of the
"smoothness hypothesis":
%
%Let us suppose that  --- due to some external reasons --- one
%cannot use the partial wave expansions for $A^\pm$ and $B^\pm$.
%Instead, it is possible to attract the "smoothness hypothesis":
the amplitudes $A^\pm$ and $B^\pm$  in the low energy region area
are smooth functions of their variables $\nu,t$:
%, these
%functions being of the form:
\begin{eqnarray}
A^+(\nu,t)=\alpha_0+\alpha_1\cdot t \  & &
A^-(\nu,t)=8a\nu \nonumber \\
B^-(\nu,t)=2(\beta_0+\beta_1\cdot t) \ & &
B^+(\nu,t)=4b\nu
\label{i33}
\end{eqnarray}

The expressions (\ref{i33}) are the most general linear forms
consistent with the requirements (\ref{i25}).
 These forms are determined by 6
parameters (for the sake of simplicity we consider them to be
real) which describe the cross section of each channel (\ref{i26}).

Let us consider now 4 instructive examples I,...,IV for
the sensitive interplay of the six parameters.
%corresponding to some special values of the latter parameters.
In each case we write down explicitly:
%\begin{description}
%\item[a. ]
the values of the parameters (a), the
%\item[b. ] The
exact expressions for $|T_r|^2$ in terms of $s,t$ (b) and
%\item[c. ] The
the "extrapolation laws" for the total cross
sections $\sigma_r\sim\int |T_r|^2 dt$, i.e. the leading order
terms of the low energy expansion of $\sigma_r$ in power series
of CMS momentum $k^2$ (c), together with short comments (d).
%\item[d.] Short comments.
%\end{description}

\noindent
\underline{Example I}\\
\makebox[1cm][l]{a.} $\alpha_0=\alpha_1=a=b=\beta_1=0, \ \ \beta_0=d$ \\
\makebox[1cm][l]{b.} $|T_1|^2=|T_2|^2=\frac12
|T_3|^2 =d^2\{(s-2m^2)^2+(s-m^2)t\}$\\
\makebox[1cm][l]{c.} $\sigma_1\simeq\sigma_2\simeq\frac12\sigma_3\simeq
k^0(1+O(k^2))$ \\
\makebox[1cm][l]{d.} The angular distributions are relatively
simple (linear in $t$) in all channels, moreover, they are
identical due to the presence of the dynamical symmetry (i.e.
conditioned solely by the particular values of the parameters).
The form of the extrapolation law in each channel is
"natural" ($\sim k^0$)

\noindent
\underline{Example II}\\
\makebox[1cm][l]{a.} $\alpha_0=\alpha_1=b=\beta_0=\beta_1=0, \ \ a=d/4m^2$ \\
\makebox[1cm][l]{b.} $|T_1|^2=|T_2|^2=\frac12 |T_3|^2 = $\\
\makebox[1cm][l]{} $d^2/16m^2\{16m^2(s-2m^2)^2-4(s-2m^2)(s-6m^2)t-
4(s-3m^2)t^2-t^3\}$\\
\makebox[1cm][l]{c.} $\sigma_1\simeq\sigma_2\simeq\frac12\sigma_3\simeq
k^0(1+O(k^2))$ \\
\makebox[1cm][l]{d.} The same situation as in I with the only
difference, that the angular distributions are more complicated.

\noindent
\underline{Example III}\\
\makebox[1cm][l]{a.} $\alpha_1=a=b=\beta_1=0, \ \ \alpha_0=\beta_0=d$ \\
\makebox[1cm][l]{b.} $|T_1|^2= d^2\{(s-4m^2)^2+(s-4m^2)t\}$ \\
\makebox[1cm][l]{} $|T_2|^2=d^2\{s^2+st\}$  \\
\makebox[1cm][l]{}  $|T_3|^2= 2d^2\{(s-2m^2)^2+(s-m^2)t\}$\\
\makebox[1cm][l]{c.} $\sigma_1\simeq k^4
\ll \sigma_2\simeq2\sigma_3\simeq  2m^4 (1+O(k^2))$\\
\makebox[1cm][l]{d.} Simple angular distributions (linear in
$t$) in all channels. The "unnatural" form of the extrapolation
law ($\sim k^4$) in the first channel, opposite to the natural forms
in the third and second channels. In addition here is a dynamical
symmetry relation between the total cross sections $\sigma_2$
and $\sigma_3$.

\noindent
\underline{Example IV}\\
\makebox[1cm][l]{a.} $\alpha_1=a=b=\beta_1=d/16m^2,
\ \ \alpha_0=\beta_0=-d/4$ \\
\makebox[1cm][l]{b.}
$|T_1|^2= d^2/(16m^2)^2\cdot 4s^2
 \{(s-4m^2)^2+(s-4m^2)t\}$\\
\makebox[1cm][l]{} $|T_2|^2=d^2/(16m^2)^2\cdot 4s\{(s-4m^2)^2+
(s-4m^2)(3s-4m^2)t$  \\
\makebox[1cm][l]{}$+(3s-8m^2)t^2+t^3\}$ \\
\makebox[1cm][l]{}  $|T_3|^2= d^2/(16m^2)^2\cdot 2 \{
-4m^2s(s-4m^2)t+s(s-8m^2)t^2-m^2t^3\}$\\
\makebox[1cm][l]{c.} $\sigma_1\simeq 3/2\sigma_2\simeq
6\sigma_3\simeq  k^4$\\
\makebox[1cm][l]{d.} The angular distributions in three
channels are quite different, the term $\sim t^0$ being absent in
the third channel. Unnatural forms ($\sim k^4$) of the
extrapolation law show up in all channels.

\begin{center}
{\large \bf Table Captions}
\end{center}

\noindent
\underline{Table 1:}\\
List of all data
for the total cross section from threshold
up to $k_1^{lab} = 500~MeV/c$ for various $\pi \pi N$ final states.\\
a)~The $\pi^+\pi^-n$\,--\,channel\\
b)~The $\pi^-\pi^0p$\,--\,channel\\
c)~The $\pi^0\pi^0n$\,--\,channel\\
d)~The $\pi^+\pi^0p$\,--\,channel\\
e)~The $\pi^+\pi^+n$\,--\,channel\\[0.5cm]

\noindent
\underline{Table 2:}\\
Basic channels of the $\pi N\rightarrow\pi\pi N$ reaction.
The isospin amplitudes $A,B,C,D$
are defined in
eq. (\protect \ref{Amp}).\\[0.5cm]

\noindent
\underline{Table 3:}\\
The results of the first step of the data filtration. The
         different columnes denote: \protect \\
         Column 1:~the kind of the fit.\protect \\
         Column 2:~the total value of $\chi^2$ corresponding to
         one  of the best solutions.\protect \\
         Column 3:~the value of $\bar\chi^2$
                   (from eq. (\protect \ref{chisquare})) \protect \\
         Column 4:~the list of the "doubtful points" (DP's);
         (i.e. points $P_q$ with $\chi^2_q\geq 5$)
         numbers
         are given in accordance with Table~1.
         \protect \\
         Column 5:~summed contribution $\chi^2(DP)$ of all DP's
         listed in column~4 to the total value of $\chi^2$.\\[0.5cm]

\noindent
\underline{Table 4:}\\
The results of the second step of the data filtration.
The notation is as in Table 3, the numbers in bracket refer to points
with a $\chi^2$ close to 5.\\[0.5cm]

\noindent
\underline{Table 5:}\\
Comparison of the $\chi^2$ for selected data points from different
$\pi \pi N$ channels in the analysis from Burkhardt and Lowe \protect
\cite{18,68} with the results of this analysis.\\[0.5cm]

\noindent
\underline{Table 6:}\\
Examples of solutions obtained in a course of the fitting with
different groups of parameters.\\

\newpage

%*******************************************************************
%Table 1

\begin{table}[h]
\caption{List of all data
for the total cross section from threshold
up to $k_1^{lab} = 500~MeV/c$ for various $\pi \pi N$ final states.
%Besides the experimental errors (4th column) the error for $\tilde M-i$
%(6th column) was calculated using eq. \protect \ref{mmgl10}
\protect \newline
a)~The $\pi^+\pi^-n$\,--\,channel}
%The values are given in mbarn.}
\begin{center}
\begin{tabular}{|l|c|l|l|c|c|c|}
\hline
\vspace{0,5mm}
     {channel} &
     {$k_1^{lab}$ in $MeV/c$} &
     {$\sigma_{tot}$ in $mb$} &
     {error} &
     {$\tilde M_i$} &
     {error} &
     {reference}\\
\hline
\hline
$\pi^+ \pi^- n$ & & & & & & \\
\hline
1&      295 &   0.0051 &  0.0012& 862  &101 &  \cite{6}  \\
2&      313 &   0.0138 &  0.0015& 685  &37  &  \cite{54} \\
3&      315 &   0.020  &  0.0031& 781  &60  &  \cite{6}  \\
4&      321 &   0.015  &  0.003 & 584  &58  &  \cite{42} \\
5&      334 &   0.051  &  0.012 & 837  &99  &  \cite{6}  \\
6&      334 &   0.027  &  0.005 & 609  &56  &  \cite{42} \\
7&      336 &   0.031  &  0.02  & 624  &208 &  \cite{55} \\
8&      342 &   0.0603 &  0.0032& 801  &21  &  \cite{54} \\
9&      346 &   0.053  &  0.013 & 712  &87  &  \cite{42} \\
10&     354 &   0.118  &  0.020 & 959  &81  &  \cite{6}  \\
11&     360 &   0.125  &  0.028 & 922  &103 &  \cite{42} \\
12&     361 &   0.060  &  0.015 & 638  &80  &  \cite{56} \\
13&     369 &   0.166  &  0.0061& 965  &18  &  \cite{54} \\
14&     374 &   0.211  &  0.036 & 1039  &89  & \cite{6}  \\
15&     374 &   0.14   &  0.10  & 843  &301 &  \cite{57} \\
16&     378 &   0.16   &  0.06  & 869  &163 &  \cite{42} \\
17&     392 &   0.4    &  0.2   & 1229  &307 & \cite{39} \\
18&     394 &   0.327  &  0.041 & 1095 &69  &  \cite{6}  \\
19&     395 &   0.374  &  0.0146& 1161 &23  &  \cite{54} \\
20&     404 &   0.38   &  0.09  & 1098 &130 &  \cite{42} \\
21&     408 &   0.546  &  0.0306& 1279 &36  &  \cite{54} \\
22&     413 &   0.477  &  0.056 & 1163 &68  &  \cite{6}  \\
23&     415 &   0.57   &  0.06  & 1253 &66  &  \cite{48} \\
24&     428 &   0.71   &  0.17  & 1333 &88  &  \cite{57} \\
25&     432 &   0.785  &  0.104 & 1249 &150 &  \cite{6}  \\
26&     449 &   1.160  &  0.052 & 1483 &33  &  \cite{54} \\
27&     451 &   1.052  &  0.125 & 1409 &84  &  \cite{6}  \\
28&     456 &   1.0    &  0.2   & 1338 &134 &  \cite{44} \\
29&     458 &   1.39   &  0.05  & 1562 &28  &  \cite{43} \\
30&     477 &   1.880  &  0.077 & 1677 &34  &  \cite{54} \\
31&     485 &   1.93   &  0.16  & 1645 &68  &  \cite{40} \\
32&     485 &   2.4    &  0.26  & 1834 &99  &  \cite{41} \\
33&     491 &   1.93   &  0.37  & 1608 &154 &  \cite{57} \\
34&     495 &   2.6    &  0.26  & 1840 &92  &  \cite{41} \\
35&     499 &   2.40   &  0.16  & 1740 &58  &  \cite{44} \\
\hline
\end{tabular}
\end{center}
\end{table}

\newpage
\clearpage
\addtocounter{table}{-1}
\begin{table}[h]
\caption{b)~The $\pi^-\pi^0 p$\,--\,channel}
\begin{center}
\begin{tabular}{|l|c|l|l|c|c|c|}
\hline
channel & $k_1^{lab}$ in $MeV/c$ &
     $\sigma_{tot}$ in $mb$ & error &
     {$\tilde M_i$}
     &error& reference\\
\hline
$\pi^- \pi^o p$ & & & & & & \\
\hline
\hline
36&     295  &  0.00075 &  0.0004& 216       & 58        & \cite{7}  \\
37&     315  &  0.0022  &  0.0007& 210       & 33        & \cite{7}  \\
38&     334  &  0.0085  &  0.0016& 301       & 28        & \cite{7}  \\
39&     354  &  0.020   &  0.005 & 356       & 44        & \cite{7}  \\
40&     375  &  0.027   &  0.006 & 342       & 38        & \cite{7}  \\
41&    391   &  0.08    &  0.13  & 517       & 420       & \cite{39} \\
42&     394  &  0.050   &  0.013 & 400       & 52        & \cite{7}  \\
43&     406  &  0.2     &   0.1  & 738       & 184       & \cite{37} \\
44&     413  &  0.073   &  0.015 & 428       & 44        & \cite{7}  \\
45&     415  &  0.11    &  0.05  & 520       & 118       & \cite{49} \\
46&     415  &  0.09    &  0.01  & 469       & 26        & \cite{48} \\
47&     428  &  0.14    &  0.07  & 526       & 142       & \cite{41} \\
48&     432  &  0.119   &  0.020 & 493       & 41        & \cite{7}  \\
49&     450  &  0.157   &  0.037 & 520       & 61        & \cite{7}  \\
50&     456  &  0.17    &  0.05  & 527       & 78        & \cite{44} \\
51&     458  &  0.17    &  0.01  & 523       & 15        & \cite{43} \\
52&     495  &  0.31    &  0.07  & 613       & 69        & \cite{41} \\
53&     499  &  0.32    &  0.05  & 613       & 48        & \cite{44} \\
\hline
\end{tabular}
\end{center}
\end{table}
\newpage

\addtocounter{table}{-1}
\begin{table}[h]
\caption{c)~The $\pi^0\pi^0n$\,--\,channel}
\vspace{0.5cm}
\begin{center}
\begin{tabular}{|l|c|l|l|c|c|c|}
\hline
channel & $k_1^{lab}$ in $MeV/c$ &
     $\sigma_{tot}$ in $mb$ & error &
     {$\tilde M_i$}
     &error& reference\\
\hline
$\pi^o \pi^o n$ & & & & & & \\
\hline
\hline
54&     272  & 0.00038 & 0.00010& 535   & 70   &  \cite{17} \\
55&     276  & 0.00059 & 0.00014& 466   & 55   &  \cite{17} \\
56&     280  & 0.00118 & 0.00022& 465   & 43   &  \cite{17} \\
57&     284  & 0.00206 & 0.00035& 477   & 40   &  \cite{17} \\
58&     286  & 0.00231 & 0.00065& 461   & 65   &  \cite{17} \\
59&     287  & 0.00333 & 0.00064& 523   & 50   &  \cite{17} \\
60&     291  & 0.00381 & 0.00081& 471   & 50   &  \cite{17} \\
61&     293  & 0.0081  & 0.0013 & 648   & 52   &  \cite{17} \\
62&     298  & 0.0085  & 0.0010 & 563   & 33   &  \cite{17} \\
63&     305  & 0.0171  & 0.0019 & 661   & 37   &  \cite{17} \\
64&     310  & 0.032   & 0.005  & 810   & 63   &  \cite{53} \\
65&     314  & 0.0219  & 0.0020 & 615   & 28   &  \cite{17} \\
66&     323  & 0.0303  & 0.0030 & 620   & 31   &  \cite{17} \\
67&     331  & 0.0598  & 0.0064 & 772   & 41   &  \cite{17} \\
68&     339  & 0.0752  & 0.0073 & 770   & 37   &  \cite{17} \\
69&     349  & 0.0981  & 0.0093 & 785   & 37   &  \cite{17} \\
70&     353  & 0.13    & 0.02   & 870   & 67   &  \cite{53} \\
71&     359  & 0.118   & 0.011  & 781   & 36   &  \cite{17} \\
72&     385  & 0.32    & 0.04   & 1040  & 65   &  \cite{51} \\
73&     390  & 0.388   & 0.046  & 1108  & 66   &  \cite{17} \\
74&     391  & 0.27    & 0.07   & 913   & 118  &  \cite{47} \\
75&     400  & 0.479   & 0.049  & 1151  & 59   &  \cite{17} \\
76&     494  & 1.3     & 0.1    & 1236  & 48   &  \cite{41} \\
\hline
\end{tabular}
\end{center}
\end{table}
\newpage

\addtocounter{table}{-1}
\begin{table}[h]
\caption{d)~The $\pi^+\pi^0p$\,--\,channel}
\vspace{0.5cm}
\begin{center}
\begin{tabular}{|l|c|l|l|c|c|c|}
\hline
channel & $k_1^{lab}$ in $MeV/c$ &
     $\sigma_{tot}$ in $mb$ & error &
     {$\tilde M_i$}
     &error& reference\\
\hline
$\pi^+ \pi^o p$ & & & & & & \\
\hline
\hline
77&     298 &   0.001  &  0.0017 & 223 & 190    &  \cite{25} \\
78&     309 &   0.0027 &  0.0012 & 267 & 59     &  \cite{25} \\
79&     331 &   0.0068 &  0.0018 & 278 & 37     &  \cite{25} \\
80&     342 &   0.018  &  0.012  & 387 & 129    &  \cite{50} \\
81&     352 &   0.0146 &  0.0026 & 309 & 28     &  \cite{25} \\
82&     374 &   0.026  &  0.0027 & 336 & 17     &  \cite{25} \\
83&     392 &   0.048  &  0.034  & 399 & 141    &  \cite{50} \\
84&     415 &   0.12   &  0.05   & 542 & 113    &  \cite{49} \\
85&     418 &   0.189  &  0.036  & 669 & 64     &  \cite{9}  \\
\hline
\end{tabular}
\end{center}
\end{table}

\newpage

\addtocounter{table}{-1}
\begin{table}[h]
\caption{e)~The $\pi^+\pi^+n$\,--\,channel}
\vspace{0.5cm}
\begin{center}
\begin{tabular}{|l|c|l|l|c|c|c|}
\hline
channel & $k_1^{lab}$ in $MeV/c$ &
     $\sigma_{tot}$ in $mb$ & error &
     {$\tilde M_i$}
     &error& reference\\
\hline
\hline
$\pi^+ \pi^+ n$ & & & & & & \\
\hline
86&     288  & 0.00011 & 0.00004 & 236 & 32   &  \cite{16} \\
87&     292  & 0.00028 & 0.00005 & 249 & 22   &  \cite{16} \\
88&     298  & 0.0018  & 0.0004  & 444 & 49   &  \cite{8}  \\
89&     299  & 0.00060 & 0.00010 & 243 & 20   &  \cite{16} \\
90&     310  & 0.00146 & 0.00022 & 246 & 19   &  \cite{16} \\
91&     317  & 0.0080  & 0.0018  & 467 & 53   &  \cite{8}  \\
92&     335  & 0.0094  & 0.0023  & 338 & 41   &  \cite{52} \\
93&     338  & 0.0217  & 0.0045  & 513 & 53   &  \cite{8}  \\
94&     342  & 0.030   & 0.018   & 565 & 170  &  \cite{50} \\
95&     358  & 0.0274  & 0.0052  & 440 & 42   &  \cite{8}  \\
96&     364  & 0.0228  & 0.0048  & 378 & 40   &  \cite{52} \\
97&     378  & 0.039   & 0.007   & 432 & 39   &  \cite{8}  \\
98&     390  & 0.026   & 0.055   & 317 & 336  &  \cite{50} \\
99&     398  & 0.0451  & 0.0103  & 395 & 45   &  \cite{8}  \\
100&    418  & 0.065   & 0.0135  & 415 & 43   &  \cite{8}  \\
101&    430  & 0.0537  & 0.010   & 353 & 33   &  \cite{52} \\
102&    439  & 0.074   & 0.0153  & 394 & 41   &  \cite{8}  \\
103&    459  & 0.083   & 0.0178  & 380 & 41   &  \cite{8}  \\
104&    477  & 0.12    & 0.01    & 424 & 18   &  \cite{38} \\
105&    480  & 0.094   & 0.0200  & 370 & 39   &  \cite{8}  \\
\hline
\end{tabular}
\end{center}
\end{table}
\clearpage

%******************************************************
%Table2

\begin{table}[h]
\caption{Basic channels of the $\pi N\rightarrow\pi\pi N$ reaction.
The isospin amplitudes $A,B,C,D$
 are defined in
eq. (\protect \ref{Amp}). }
\vspace{0.5cm}
\begin{center}
\begin{tabular}{||l|c|c|c||}
\hline \hline
channel & $N^0$  & amplitude &stat.factor $f$\\
\hline
$\pi^- p\rightarrow\pi^- \pi^+ n$ & 1 &$X_1=\frac{\sqrt{2}}{2}(A+C)$ &1 \\
$\pi^- p\rightarrow\pi^- \pi^0 p$ & 2 &$X_2=\frac12C-D$ &1 \\
$\pi^- p\rightarrow\pi^0 \pi^0 p$ & 3 &$X_3=\frac{\sqrt{2}}{2}A$ &1/2 \\
$\pi^+ p\rightarrow\pi^+ \pi^0 p$ & 4 &$X_4=\frac12C+D$ &1 \\
$\pi^+ p\rightarrow\pi^+\pi^+ n$  & 5 &$X_1=\frac{\sqrt{2}}{2}(B+C)$ &1/2 \\
\hline \hline
\end{tabular}
\end{center}
\end{table}

\vspace{1cm}

%*********************************************************************
%Table 3

\begin{table}[h]
\caption{The results of the first step of the data filtration. The
         different columnes denote: \protect \\
         Column 1:~the kind of the fit.\protect \\
         Column 2:~the total value of $\chi^2$ corresponding to
         one  of the best solutions.\protect \\
         Column 3:~the value of $\bar\chi^2$
                   (from eq. (\protect \ref{chisquare})) \protect \\
         Column 4:~the list of the "doubtful points" (DP's);
         (i.e. points
         $P_q$ with $\chi^2_q\geq 5$);
         numbers
         are given in accordance with Table~1 \protect \\
         Column 5:~summed contribution $\chi^2(DP)$ of all DP's
         listed in column~4 to the total value of $\chi^2$.}
\vspace{0.5cm}
\begin{center}
\begin{tabular}{||l|c|c|l|c||}
\hline \hline
Fit & $\chi^2$  &$\bar\chi^2$&List of DP's& $\chi^2(DP)$ \\
\hline
$B$           & 188.6  &2.03& 4,6,64,73,75,76,88,91,93,95 & 87.6   \\
$\pi$         & 284.7  &2.85& many                        & ---   \\
$\Delta$      & 1396.0 &14.2& many                        &  ---  \\
$B+\pi$       & 149.0  &1.66& 1,6,64,73,75,88,90,91,93    &  64.6  \\
$B+\Delta$    & 180.8  &2.08& 4,6,64,73,75,76,88,91,93    &  78.1  \\
$\pi+\Delta$  & 204.4  &2.17& 4,6,64,67,72,73,75,76,85
,88,90,91,93,95&  109.7                                            \\
$B+\pi+\Delta$& 148.2  &1.79&  1,6,64,73,75,88,90,91,93   & 64.1  \\
\hline \hline
\end{tabular}
\end{center}
\end{table}

\newpage
\clearpage

%**********************************************************************
%Table4

\begin{table}[h]
\caption{The results of the second step of the data filtration.
The notation is as in Table 3, the numbers in bracket refer to points
with a $\chi^2$ close to 5.}
\vspace{0.5cm}
\begin{center}
\begin{tabular}{||l|c|c|l|c||}
\hline \hline
Fit & $\chi^2$  &$\bar\chi^2$&List of DP's& $\chi^2(DP)$ \\
\hline
$B$           & 123.4  &1.43& [1]4,67,72,76,95            & 30.6   \\
$\pi$         & 216.3  &2.33& many                        & ---   \\
$\Delta$      & 1259.0 &13.8& many                        &  ---  \\
$B+\pi$       & 98.8   &1.20& 1,[4],90                    &  11.8  \\
$B+\Delta$    & 119.1  &1.49& [1],4,67,72                 &  17.3  \\
$\pi+\Delta$  & 139.5  &1.60& 4,19,21,67,72,85,95,97      &  52.6 \\
$B+\pi+\Delta$& 98.2   &1.29&  1,[4],85                   & 12.0  \\
\hline \hline
\end{tabular}
\end{center}
\end{table}

\vspace{1cm}

%****************************************************************
%Table 5

\begin{table}[h]
\caption{Comparison of the $\chi^2$ for
selected data points from different
$\pi \pi N$ channels in the analysis from Burkhardt and Lowe \protect
\cite{18,68} with the results of this analysis.}
\vspace{0.5cm}
\begin{center}
\begin{tabular}{||l|c|c||}
\hline \hline
$no.$ point  & \multicolumn{2}{c||}{Individual contribution $\chi^2_i$} \\
\cline{2-3}
(channel)  &\hspace{0.5cm} BL\hspace{0.5cm}   & Our results \\ \hline
1 ($\pi^-\pi^+n$)    & 5.8  & 5.8         \\
4 ($\pi^-\pi^+n$)    & 4.2  & 3.9         \\
6 ($\pi^-\pi^+n$)    & 9.6  & 9.1         \\ \hline
64 ($\pi^0\pi^0n$)   & 8.1  & 8.4         \\
73 ($\pi^0\pi^0n$)   & 4.4  & 3.7          \\
75 ($\pi^0\pi^0n$)   & 5.3  & 3.5          \\ \hline
88 ($\pi^+\pi^+n$)   & 10.0 & 9.0          \\
91 ($\pi^+\pi^+n$)   & 8.9  & 7.4           \\
93 ($\pi^+\pi^+n$)   & 10.4 & 7.7            \\
\hline \hline
\end{tabular}
\end{center}
\end{table}

%****************************************************
%Table 6

\begin{table}[h]
\caption{Examples of solutions obtained in a course of the fitting with
different groups of parameters }
\vspace{0.5cm}
\begin{center}
\begin{tabular}{||c||c|c|c||c|c|c||c|c|c||}
\hline
\multicolumn{4}{|c|}{$B(\bar\chi^2=1.4)$} &
\multicolumn{3}{c|}{$\pi(\bar\chi^2=2.3)$} &
\multicolumn{3}{c|}{$B+\pi(\bar\chi^2=2.3)$}\\ \hline
Para- &value  & error & correlation &
value  &error &correlation &
value  &error &correlation \\
meter &$ 10^3$ &$ 10^3$ & factor $ 10^3$ &
 &        & factor  $ 10^3$ &
 $ 10^3$ & $ 10^4$ & factor  $ 10^4$ \\
\hline
\hline
 $A_1$   &  0.39 & 1.12 & 4.48 &  0.0 & fixed & - & -2.70 & 0.18 & 0.20 \\
  $ A_2$ & -5.00 & 10.9 & 6.79 &  0.0 & fixed & - & -4.46 & 4.58 & 2.37 \\
  $ A_3$ & -10.4 & 36.7 & 27.2 &  0.0 & fixed & - & -12.9 & 7.56 & 2.87  \\
  $ A_4$ & -0.77 & 0.35 & 2.11 &  0.0 & fixed & - & -2.10 & 0.34 & 10.6 \\
  $ A_5$ &  3.77 & 11.1 & 21.2 &  0.0 & fixed & - & -4.43 & 6.59 & 21.8 \\
  $ A_6$ &  0.76 & 6.09 & 25.8 &  0.0 & fixed & - &  5.63 & 0.34 & 0.30 \\
  $ A_7$ & -6.11 & 7.67 & 4.43 &  0.0 & fixed & - &  -3.16& 11.6 & 48.9 \\
  $ A_8$ &  0.85 & 1.86 & 3.66 &  0.0 & fixed & - & 1.84  & 1.81 & 17.9 \\
  $ A_9$ &  6.34 & 6.19 & 1.09 &  0.0 & fixed & - & 7.88  & 1.13 & 0.39 \\
  $ A_10$&  0.41 & 0.55 & 0.26 &  0.0 & fixed & - & -0.15 & 0.11 & 0.06 \\
  $ A_11$&  8.86 & 6.33 & 0.44 &  0.0 & fixed & - &  0.37 & 2.14 & 0.40 \\
  $ G_0 $&  0.00 & fixed&  -   &$9.7$ & 3.53  & 4.7&-0.18 & 0.03 & 0.57 \\
  $ G_1 $&  0.00 & fixed&  -   &$-672$ & 44.4  &3.1& -2.74 & 0.17 & 0.12 \\
  $ G_2 $&  0.00 & fixed&  -   &$1.5\cdot10^4$&427&4.7&  20.4 & 12.2 &
  3.36 \\
  $ G_3 $&  0.00 & fixed&  -   &$2.9\cdot 10^4$& 405   &3.1&  18.7 & 11.3
   & 2.56 \\
   \hline \end{tabular}
   \end{center}
   \end{table}

\begin{center}
{\large \bf Figure Captions}
\end{center}

\noindent
\underline{Figure 1:}\\
Schematical decomposition of the $\pi N \to \pi \pi N$ reaction
amplitude into the background part ($B$), the $\Delta$-isobar piece ($\Delta$)
and the OPE-term ($\Pi$).\\[1cm]

\noindent
\underline{Figure 2:}\\
Different contributions of the $\Delta$-isobar.
(a), (b), (c): $\Delta$-excitation in the first, second and both
intermediate states, respectively; (d): nonlinear $\pi \pi N \Delta$
coupling.\\[1cm]

\noindent
\underline{Figure 3:}\\
Comparison of the model independent fit to the quasi amplitude $\tilde M_i$
(eq. (\protect \ref{MMgl10})) in 5 $\pi \pi N$  channels with the data
as a function of $p_{\pi}\,(GeV/c)$. Note that ``doubtful points'',
as discussed in Sect. 3.4, are not included in the data.
The experimental data and corresponding references
are from Table 1. Data points from Kernel et al. \protect \cite{6,7,8,9}
are emphasized by asterisks.
\protect \newline
a)~The $\pi^+\pi^-n$\,--\,channel.\\
b)~The $\pi^-\pi^0p$\,--\,channel.\\
c)~The $\pi^0\pi^0n$\,--\,channel.\\
d)~The $\pi^+\pi^0p$\,--\,channel.\\
e)~The $\pi^+\pi^+n$\,--\,channel.
Compared are the fits with the full data set and with a
partial data set
(lower and
upper line, respectively, 4 points out of 20 are suppressed).

\newpage

%****************************************************************
%Figure 1

\vspace{2cm}

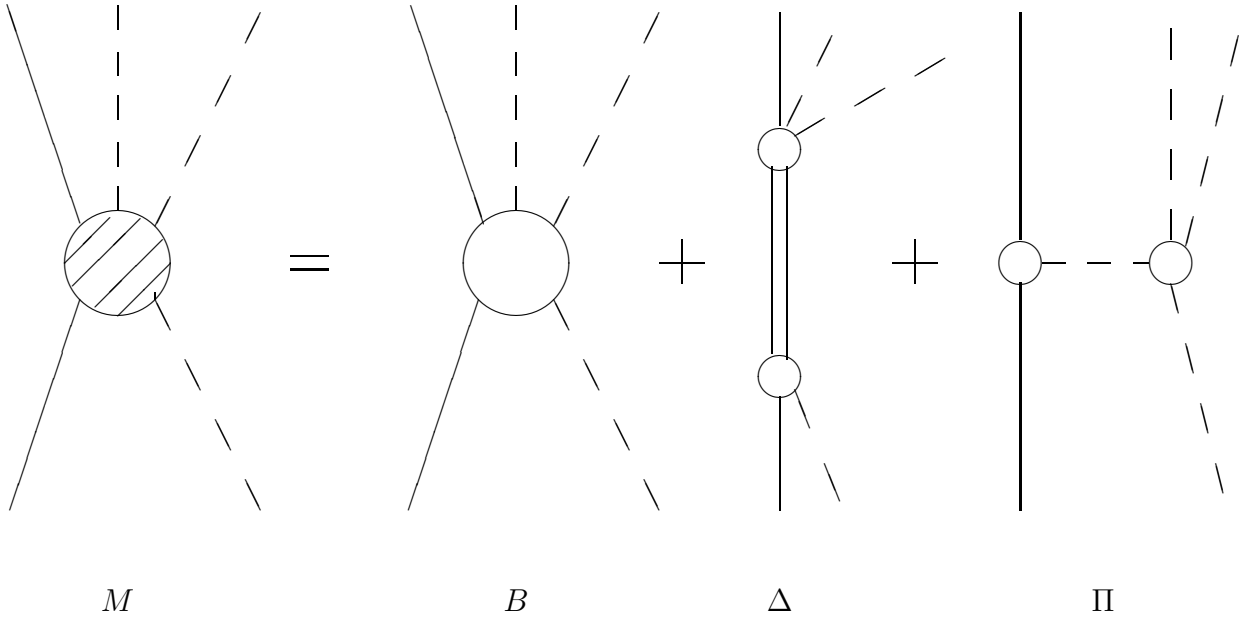
\begin{figure}[b]
\unitlength=1.00mm
%\special{em:linewidth 0.4pt}
\linethickness{0.4pt}
\begin{picture}(164.,70.)
\put(-5.,-50.){
\begin{picture}(164.00,70.)
\put(15.00,104.89){\circle{14.00}}
\put(15.00,112.00){\line(0,1){3.11}}
\put(15.00,117.78){\line(0,1){3.11}}
\put(15.00,124.00){\line(0,1){3.11}}
\put(15.00,129.78){\line(0,1){3.11}}
\put(20.00,109.78){\line(1,2){2.00}}
\put(24.00,117.78){\line(1,2){2.00}}
\put(28.00,125.78){\line(1,2){2.00}}
\put(20.00,100.89){\line(0,-1){0.89}}
\put(20.00,100.00){\line(1,-2){2.00}}
\put(24.00,92.00){\line(1,-2){2.00}}
\put(28.00,84.00){\line(1,-2){2.00}}
\put(9.00,101.78){\line(1,1){9.00}}
\put(21.00,108.00){\line(-1,-1){9.00}}
\put(15.00,97.78){\line(1,1){7.00}}
\put(8.00,104.89){\line(1,1){6.00}}
\put(38.00,104.00){\line(1,0){5.00}}
\put(43.00,105.78){\line(-1,0){5.00}}
\put(68.00,112.00){\line(0,1){3.11}}
\put(68.00,117.78){\line(0,1){3.11}}
\put(68.00,124.00){\line(0,1){3.11}}
\put(68.00,129.78){\line(0,1){3.11}}
\put(73.00,109.78){\line(1,2){2.00}}
\put(77.00,117.78){\line(1,2){2.00}}
\put(81.00,125.78){\line(1,2){2.00}}
\put(73.00,100.00){\line(1,-2){2.00}}
\put(77.00,92.00){\line(1,-2){2.00}}
\put(81.00,84.00){\line(1,-2){2.00}}
\put(68.00,104.89){\circle{14.00}}
\put(103.00,120.00){\circle{6.00}}
\put(103.00,89.78){\circle{6.00}}
\put(102.00,92.89){\line(0,1){24.89}}
\put(104.00,117.78){\line(0,-1){25.78}}
\put(104.00,92.00){\line(0,1){25.78}}
\put(104.00,117.78){\line(0,-1){25.78}}
\put(103.00,123.11){\line(0,1){12.00}}
\put(105.00,121.78){\line(5,3){4.00}}
\put(113.00,125.78){\line(5,3){4.00}}
\put(121.00,129.78){\line(5,3){4.00}}
\put(104.00,123.11){\line(1,2){2.00}}
\put(108.00,131.11){\line(1,2){2.00}}
\put(103.00,87.11){\line(0,-1){12.00}}
\put(135.00,104.89){\circle{6.00}}
\put(155.00,104.89){\circle{6.00}}
\put(138.00,104.89){\line(1,0){3.00}}
\put(144.00,104.89){\line(1,0){3.00}}
\put(150.00,104.89){\line(1,0){2.00}}
\put(135.00,108.00){\line(0,1){27.11}}
\put(135.00,102.22){\line(0,-1){27.11}}
\put(155.00,102.22){\line(1,-4){1.00}}
\put(157.00,94.22){\line(1,-4){1.00}}
\put(159.00,86.22){\line(1,-4){1.00}}
\put(161.00,78.22){\line(1,-4){1.00}}
\put(155.00,108.00){\line(0,1){4.00}}
\put(155.00,116.00){\line(0,1){4.00}}
\put(155.00,124.00){\line(0,1){4.00}}
\put(155.00,132.00){\line(0,1){4.00}}
\put(157.00,107.11){\line(1,4){1.00}}
\put(159.00,115.11){\line(1,4){1.00}}
\put(161.00,123.11){\line(1,4){1.00}}
\put(163.00,131.11){\line(1,4){1.00}}
\put(121.00,108.00){\line(0,-1){5.78}}
\put(121.00,102.22){\line(0,1){5.78}}
\put(118.00,104.89){\line(1,0){6.00}}
\put(87.00,104.89){\line(1,0){6.00}}
\put(90.00,108.00){\line(0,-1){5.78}}
\put(15.00,60.00){\makebox(0,0)[cc]{$M$}}
\put(68.00,60.00){\makebox(0,0)[cc]{$B$}}
\put(103.00,60.00){\makebox(0,0)[cc]{$\Delta$}}
\put(146.00,60.00){\makebox(0,0)[cc]{$\Pi$}}
\put(85.00,76.00){\line(1,-2){2.00}}
\put(87.00,72.00){\line(0,0){0.00}}
\put(103.00,72.00){\line(0,1){3.11}}
\put(135.00,72.00){\line(0,1){3.11}}
\put(10.00,100.00){\line(-1,-3){9.33}}
\put(63.00,100.00){\line(-1,-3){9.33}}
\put(32.00,76.00){\line(1,-2){2.00}}
\put(32.00,134.22){\line(1,2){2.00}}
\put(15.00,136.00){\line(0,1){3.11}}
\put(85.00,134.22){\line(1,2){2.00}}
\put(103.00,135.11){\line(0,1){3.11}}
\put(135.00,135.11){\line(0,1){3.11}}
\put(68.00,136.00){\line(0,1){3.11}}
\put(10.00,110.22){\line(-1,3){9.67}}
\put(54.00,139.11){\line(1,-3){9.67}}
\put(105.00,88.00){\line(2,-5){2.00}}
\put(109.00,78.22){\line(2,-5){2.00}}
\end{picture}}
\end{picture}
\caption{Schematical decomposition of the $\pi N \to \pi \pi N$ reaction
amplitude into the background part ($B$), the $\Delta$-isobar piece ($\Delta$)
and the OPE-term ($\Pi$).}
\end{figure}

\newpage

\begin{figure}
\unitlength.1mm
\begin{picture}(1400.,1800.)
\put(0.,900.){
%\special{em:linewidth 0.4pt}
\linethickness{0.4pt}
\begin{picture}(700.00,1000.00)
\put(150.00,100.0){\line(1,1){40.00}}
\put(200.00,150.0){\line(1,1){40.00}}
\put(250.00,200.0){\line(1,1){40.00}}
\put(300.00,250.0){\line(1,1){50.00}}
\put(550.00,100.0){\line(-1,1){200.00}}
\put(350.00,300.0){\line(0,1){400.00}}
\put(150.00,900.0){\line(1,-1){40.00}}
\put(200.00,850.0){\line(1,-1){40.00}}
\put(250.00,800.0){\line(1,-1){40.00}}
\put(300.00,750.0){\line(1,-1){50.00}}
\put(150.00,700.0){\line(1,-1){40.00}}
\put(200.00,650.0){\line(1,-1){40.00}}
\put(250.00,600.0){\line(1,-1){40.00}}
\put(300.00,550.0){\line(1,-1){50.00}}
\put(550.00,900.00){\line(-1,-1){200.00}}
\put(200.00,200.0){\makebox(0,0)[cc]{$\pi_{in}$}}
\put(200.00,780.0){\makebox(0,0)[cc]{$\pi^1_{out}$}}
\put(200.00,580.0){\makebox(0,0)[cc]{$\pi^2_{out}$}}
\put(500.00,200.0){\makebox(0,0)[cc]{$N_{in}$}}
\put(500.00,770.0){\makebox(0,0)[cc]{$N_{out}$}}
\put(390.00,400.00){\makebox(0,0)[cc]{$\Delta$}}
\put(390.00,600.0){\makebox(0,0)[cc]{$N$}}
\put(90.00,950.0){\makebox(0,0)[cc]{(a)}}
\end{picture}
}
\put(700.,900.){
%\special{em:linewidth 0.4pt}
\linethickness{0.4pt}
\begin{picture}(700.00,1000.00)
\put(150.00,100.0){\line(1,1){40.00}}
\put(200.00,150.0){\line(1,1){40.00}}
\put(250.00,200.0){\line(1,1){40.00}}
\put(300.00,250.0){\line(1,1){50.00}}
\put(550.00,100.0){\line(-1,1){200.00}}
\put(350.00,300.0){\line(0,1){400.00}}
\put(150.00,900.0){\line(1,-1){40.00}}
\put(200.00,850.0){\line(1,-1){40.00}}
\put(250.00,800.0){\line(1,-1){40.00}}
\put(300.00,750.0){\line(1,-1){50.00}}
\put(150.00,700.0){\line(1,-1){40.00}}
\put(200.00,650.0){\line(1,-1){40.00}}
\put(250.00,600.0){\line(1,-1){40.00}}
\put(300.00,550.0){\line(1,-1){50.00}}
\put(550.00,900.00){\line(-1,-1){200.00}}
\put(200.00,200.0){\makebox(0,0)[cc]{$\pi_{in}$}}
\put(200.00,780.0){\makebox(0,0)[cc]{$\pi^1_{out}$}}
\put(200.00,580.0){\makebox(0,0)[cc]{$\pi^2_{out}$}}
\put(500.00,200.0){\makebox(0,0)[cc]{$N_{in}$}}
\put(500.00,770.0){\makebox(0,0)[cc]{$N_{out}$}}
\put(390.00,400.00){\makebox(0,0)[cc]{$N$}}
\put(390.00,600.0){\makebox(0,0)[cc]{$\Delta$}}
\put(90.00,950.0){\makebox(0,0)[cc]{(b)}}
\end{picture}
}
\put(0.,-90.){
%\special{em:linewidth 0.4pt}
\linethickness{0.4pt}
\begin{picture}(700.00,1000.00)
\put(150.00,100.0){\line(1,1){40.00}}
\put(200.00,150.0){\line(1,1){40.00}}
\put(250.00,200.0){\line(1,1){40.00}}
\put(300.00,250.0){\line(1,1){50.00}}
\put(550.00,100.0){\line(-1,1){200.00}}
\put(350.00,300.0){\line(0,1){400.00}}
\put(150.00,900.0){\line(1,-1){40.00}}
\put(200.00,850.0){\line(1,-1){40.00}}
\put(250.00,800.0){\line(1,-1){40.00}}
\put(300.00,750.0){\line(1,-1){50.00}}
\put(150.00,700.0){\line(1,-1){40.00}}
\put(200.00,650.0){\line(1,-1){40.00}}
\put(250.00,600.0){\line(1,-1){40.00}}
\put(300.00,550.0){\line(1,-1){50.00}}
\put(550.00,900.00){\line(-1,-1){200.00}}
\put(200.00,210.0){\makebox(0,0)[cc]{$\pi_{in}$}}
\put(200.00,780.0){\makebox(0,0)[cc]{$\pi^1_{out}$}}
\put(200.00,580.0){\makebox(0,0)[cc]{$\pi^2_{out}$}}
\put(500.00,200.0){\makebox(0,0)[cc]{$N_{in}$}}
\put(500.00,770.0){\makebox(0,0)[cc]{$N_{out}$}}
\put(390.00,400.00){\makebox(0,0)[cc]{$\Delta$}}
\put(390.00,600.0){\makebox(0,0)[cc]{$\Delta$}}
\put(90.00,950.0){\makebox(0,0)[cc]{(c)}}
\end{picture}
}
\put(700.,-90.){
%\special{em:linewidth 0.4pt}
\linethickness{0.4pt}
\begin{picture}(700.00,1000.00)
\put(150.00,100.0){\line(1,1){40.00}}
\put(200.00,150.0){\line(1,1){40.00}}
\put(250.00,200.0){\line(1,1){40.00}}
\put(300.00,250.0){\line(1,1){50.00}}
\put(550.00,100.0){\line(-1,1){200.00}}
\put(350.00,300.0){\line(0,1){400.00}}
\put(150.00,900.0){\line(1,-1){40.00}}
\put(200.00,850.0){\line(1,-1){40.00}}
\put(250.00,800.0){\line(1,-1){40.00}}
\put(300.00,750.0){\line(1,-1){50.00}}
\put(150.00,800.0){\line(2,-1){40.00}}
\put(200.00,775.0){\line(2,-1){40.00}}
\put(250.00,750.0){\line(2,-1){40.00}}
\put(300.00,725.0){\line(2,-1){50.00}}
\put(550.00,900.00){\line(-1,-1){200.00}}
\put(200.00,200.0){\makebox(0,0)[cc]{$\pi_{in}$}}
\put(200.00,730.0){\makebox(0,0)[cc]{$\pi^1_{out}$}}
\put(200.00,910.0){\makebox(0,0)[cc]{$\pi^2_{out}$}}
\put(500.00,200.0){\makebox(0,0)[cc]{$N_{in}$}}
\put(500.00,770.0){\makebox(0,0)[cc]{$N_{out}$}}
\put(390.00,500.00){\makebox(0,0)[cc]{$\Delta$}}
\put(90.00,950.0){\makebox(0,0)[cc]{(d)}}
\end{picture}
}
\end{picture}
\caption{Different contributions of the $\Delta$-isobar.
(a), (b), (c): $\Delta$-excitation in the first, second and both
intermediate states, respectively; (d): direct
 $\pi \pi N \Delta$ coupling
(the crossing graphs are implied to be taken into account also).}
\end{figure}
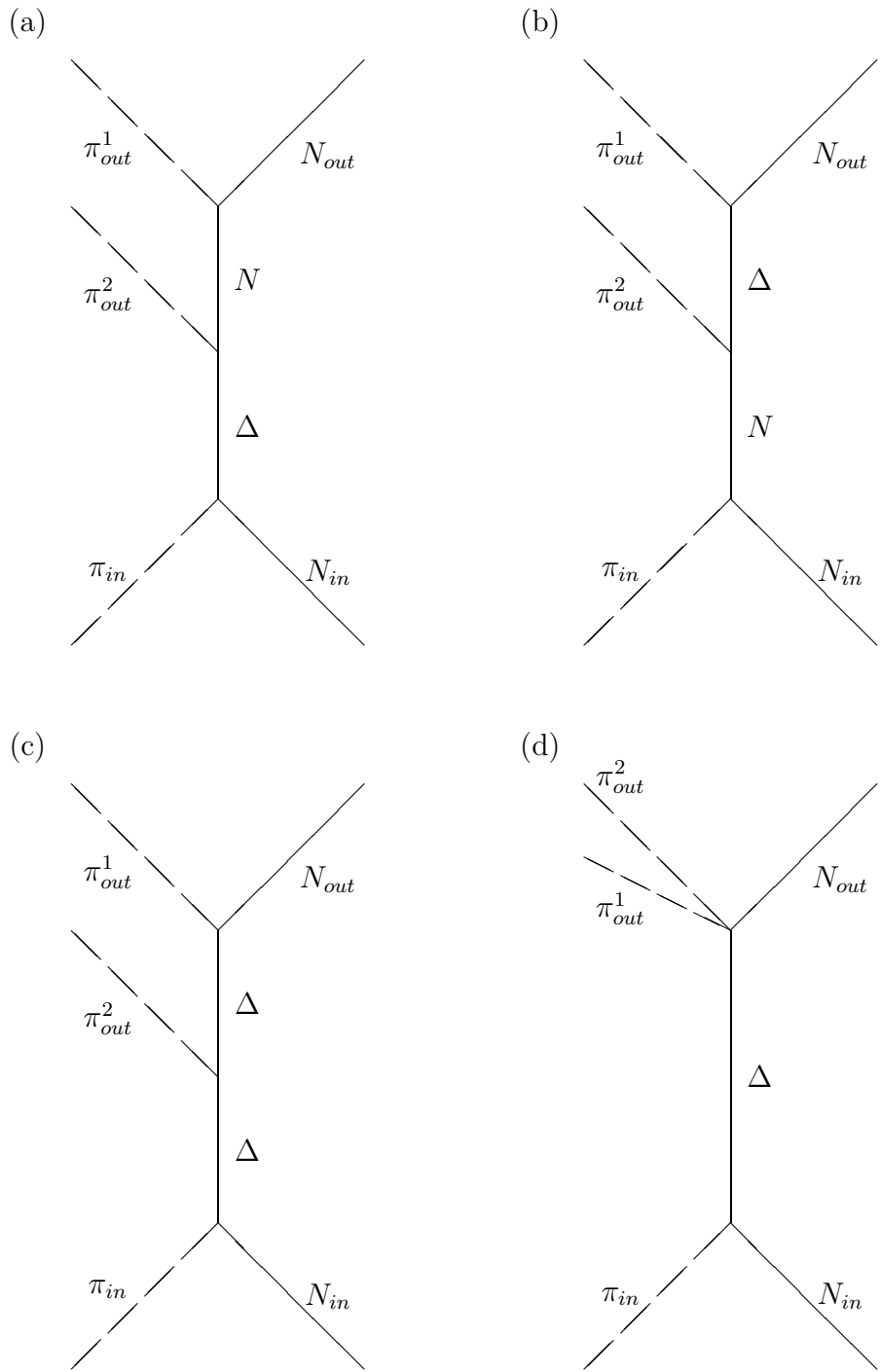

\newpage

%*******************************************************
%Figure 3

\begin{figure}
\caption
{Comparison of the model independent fit to the quasi amplitude $\tilde M_i$
(eq. (\protect \ref{MMgl10})) in 5 $\pi \pi N$  channels with the data
as a function of $p_{\pi}\,(GeV/c)$. Note that ``doubtful points'',
as discussed in Sect. 3.4, are not included in the data.
The experimental data and corresponding references
are from Table 1. Data points from Kernel et al. \protect \cite{6,7,8,9}
are emphasized by asterisks. \protect \\[5mm]
a)~The $\pi^+\pi^-n$\,--\,channel.}

\begin{minipage}{100mm}
\textheight 5mm
\textwidth 2mm
\setlength{\unitlength}{0.1mm}
\begin{picture}(1000,1332)
  \put(300,300){\framebox(1000,860)}
  \put(150,200){\makebox(300,100){0.25}}
  \put(300,299){\line(0,-1){20}}
  \put(350,200){\makebox(300,100){0.3}}
  \put(500,300){\line(0,-1){20}}
  \put(550,200){\makebox(300,100){0.35}}
  \put(700,300){\line(0,-1){20}}
  \put(750,200){\makebox(300,100){0.4}}
  \put(900,300){\line(0,-1){20}}
  \put(950,200){\makebox(300,100){0.45}}
  \put(1100,300){\line(0,-1){20}}
  \put(1150,200){\makebox(300,100){0.5}}
  \put(1300,300){\line(0,-1){20}}
  \put(-40,250){\makebox(300,100)[r]{   0}}
  \put(300,300){\line(-1,0){20}}
  \put(-40,422){\makebox(300,100)[r]{ 400}}
  \put(300,472){\line(-1,0){20}}
  \put(-40,594){\makebox(300,100)[r]{ 800}}
  \put(300,644){\line(-1,0){20}}
  \put(-40,766){\makebox(300,100)[r]{1200}}
  \put(300,816){\line(-1,0){20}}
  \put(-40,938){\makebox(300,100)[r]{1600}}
  \put(300,988){\line(-1,0){20}}
  \put(-40,1110){\makebox(300,100)[r]{2000}}
  \put(300,1162){\line(-1,0){20}}
\put(300,50){\makebox(1000,200){{\large $p_{\pi}\,\,(GeV/c)$}}}
\put(100,1080){\makebox(100,100)[l]{{\large $\tilde M_i$}}}
%\put(0,1000){\makebox(100,100)[l]{{\large $\sim$ $\sqrt{\sigma_{tot}}$}}}
\end{picture}
\end{minipage}
\end{figure}

\newpage

\addtocounter{figure}{-1}
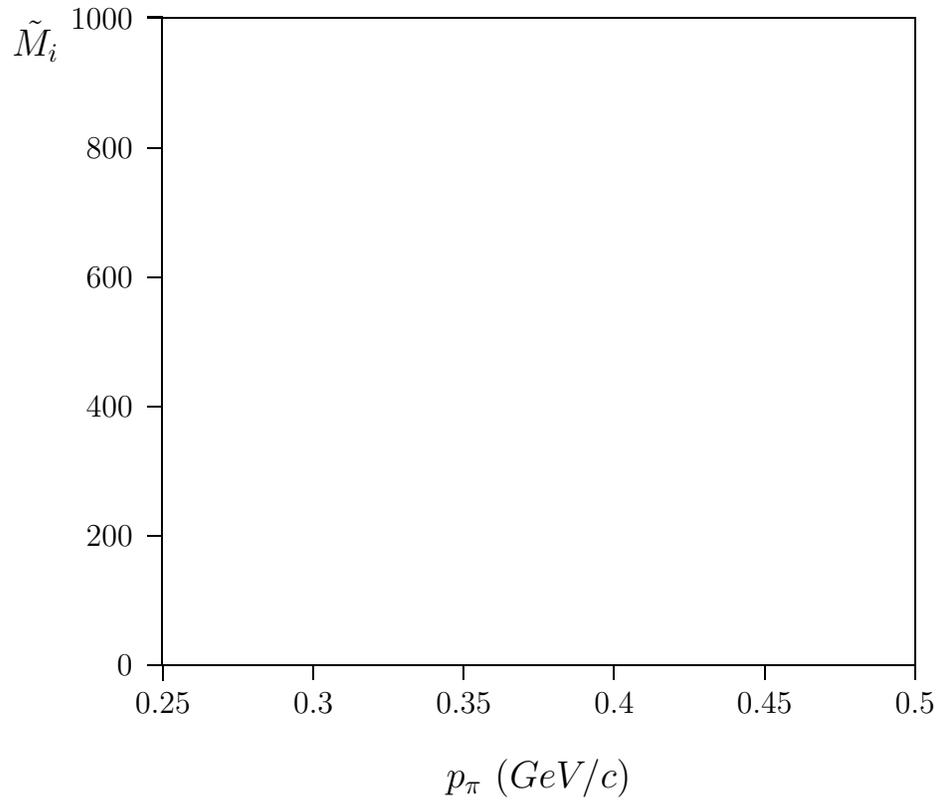
\begin{figure}[t]
\caption{b)~The $\pi^-\pi^0 p$\,--\,channel.}

\begin{minipage}{100mm}
\textheight 5mm
\textwidth 2mm
\setlength{\unitlength}{0.1mm}
\begin{picture}(1000,1332)
  \put(300,300){\framebox(1000,860)}
  \put(150,200){\makebox(300,100){0.25}}
  \put(300,299){\line(0,-1){20}}
  \put(350,200){\makebox(300,100){0.3}}
  \put(500,300){\line(0,-1){20}}
  \put(550,200){\makebox(300,100){0.35}}
  \put(700,300){\line(0,-1){20}}
  \put(750,200){\makebox(300,100){0.4}}
  \put(900,300){\line(0,-1){20}}
  \put(950,200){\makebox(300,100){0.45}}
  \put(1100,300){\line(0,-1){20}}
  \put(1150,200){\makebox(300,100){0.5}}
  \put(1300,300){\line(0,-1){20}}
  \put(-40,250){\makebox(300,100)[r]{   0}}
  \put(300,300){\line(-1,0){20}}
  \put(-40,422){\makebox(300,100)[r]{ 200}}
  \put(300,472){\line(-1,0){20}}
  \put(-40,594){\makebox(300,100)[r]{ 400}}
  \put(300,644){\line(-1,0){20}}
  \put(-40,766){\makebox(300,100)[r]{ 600}}
  \put(300,816){\line(-1,0){20}}
  \put(-40,938){\makebox(300,100)[r]{ 800}}
  \put(300,988){\line(-1,0){20}}
  \put(-40,1110){\makebox(300,100)[r]{1000}}
  \put(300,1162){\line(-1,0){20}}
\put(300,50){\makebox(1000,200){{\large $p_{\pi}\,\,(GeV/c)$}}}
\put(100,1080){\makebox(100,100)[l]{{\large $\tilde M_i$}}}
%\put(0,1000){\makebox(100,100)[l]{{\large $\sim$ $\sqrt{\sigma_{tot}}$}}}
\end{picture}
\end{minipage}
\end{figure}

\newpage

\addtocounter{figure}{-1}
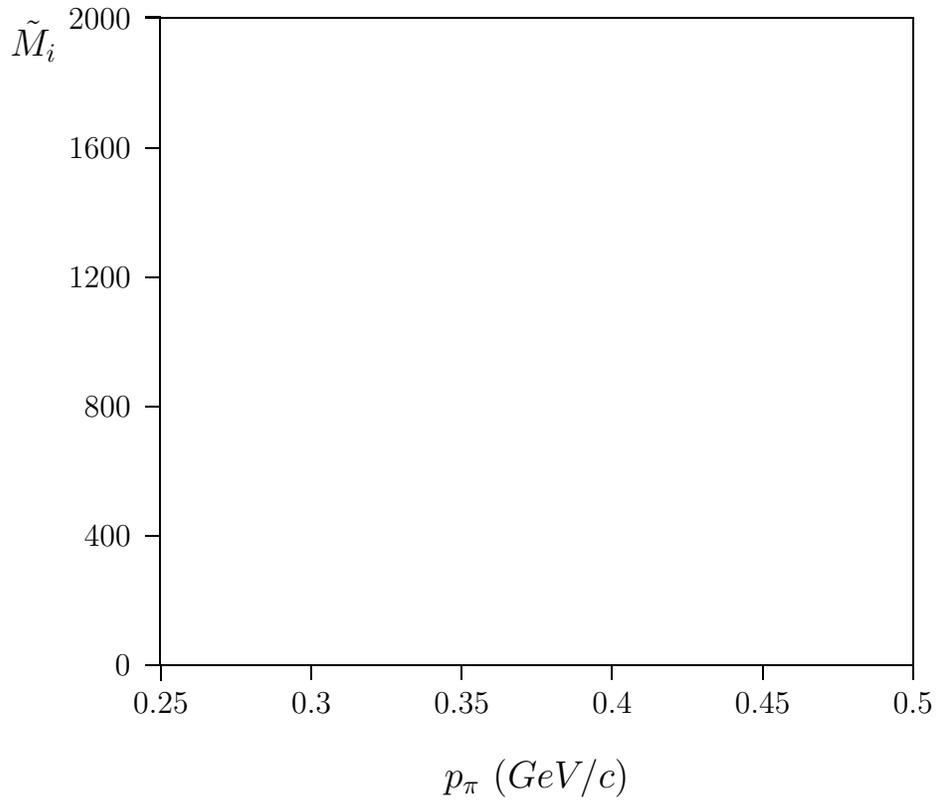
\begin{figure}[t]
\caption{c)~The $\pi^0\pi^0n$\,--\,channel.}

\begin{minipage}{100mm}
\textheight 5mm
\textwidth 2mm
\setlength{\unitlength}{0.1mm}
\begin{picture}(1000,1332)
  \put(300,300){\framebox(1000,860)}
  \put(150,200){\makebox(300,100){0.25}}
  \put(300,299){\line(0,-1){20}}
  \put(350,200){\makebox(300,100){0.3}}
  \put(500,300){\line(0,-1){20}}
  \put(550,200){\makebox(300,100){0.35}}
  \put(700,300){\line(0,-1){20}}
  \put(750,200){\makebox(300,100){0.4}}
  \put(900,300){\line(0,-1){20}}
  \put(950,200){\makebox(300,100){0.45}}
  \put(1100,300){\line(0,-1){20}}
  \put(1150,200){\makebox(300,100){0.5}}
  \put(1300,300){\line(0,-1){20}}
  \put(-40,250){\makebox(300,100)[r]{   0}}
  \put(300,300){\line(-1,0){20}}
  \put(-40,422){\makebox(300,100)[r]{ 400}}
  \put(300,472){\line(-1,0){20}}
  \put(-40,594){\makebox(300,100)[r]{ 800}}
  \put(300,644){\line(-1,0){20}}
  \put(-40,766){\makebox(300,100)[r]{1200}}
  \put(300,816){\line(-1,0){20}}
  \put(-40,938){\makebox(300,100)[r]{1600}}
  \put(300,988){\line(-1,0){20}}
  \put(-40,1110){\makebox(300,100)[r]{2000}}
  \put(300,1162){\line(-1,0){20}}
\put(300,50){\makebox(1000,200){{\large $p_{\pi}\,\,(GeV/c)$}}}
\put(100,1080){\makebox(100,100)[l]{{\large $\tilde M_i$}}}
%\put(0,1000){\makebox(100,100)[l]{{\large $\sim$ $\sqrt{\sigma_{tot}}$}}}
\end{picture}
\end{minipage}
\end{figure}

\newpage

\addtocounter{figure}{-1}
\begin{figure}[t]
\caption{d)~The $\pi^+\pi^0p$\,--\,channel.}

\begin{minipage}{100mm}
\textheight 5mm
\textwidth 2mm
\setlength{\unitlength}{0.1mm}
\begin{picture}(1000,1332)
  \put(300,300){\framebox(1000,860)}
  \put(150,200){\makebox(300,100){0.25}}
  \put(300,299){\line(0,-1){20}}
  \put(350,200){\makebox(300,100){0.3}}
  \put(500,300){\line(0,-1){20}}
  \put(550,200){\makebox(300,100){0.35}}
  \put(700,300){\line(0,-1){20}}
  \put(750,200){\makebox(300,100){0.4}}
  \put(900,300){\line(0,-1){20}}
  \put(950,200){\makebox(300,100){0.45}}
  \put(1100,300){\line(0,-1){20}}
  \put(1150,200){\makebox(300,100){0.5}}
  \put(1300,300){\line(0,-1){20}}
  \put(-40,250){\makebox(300,100)[r]{   0}}
  \put(300,300){\line(-1,0){20}}
  \put(-40,422){\makebox(300,100)[r]{ 200}}
  \put(300,472){\line(-1,0){20}}
  \put(-40,594){\makebox(300,100)[r]{ 400}}
  \put(300,644){\line(-1,0){20}}
  \put(-40,766){\makebox(300,100)[r]{ 600}}
  \put(300,816){\line(-1,0){20}}
  \put(-40,938){\makebox(300,100)[r]{ 800}}
  \put(300,988){\line(-1,0){20}}
  \put(-40,1110){\makebox(300,100)[r]{1000}}
  \put(300,1162){\line(-1,0){20}}
\put(300,50){\makebox(1000,200){{\large $p_{\pi}\,\,(GeV/c)$}}}
\put(100,1080){\makebox(100,100)[l]{{\large $\tilde M_i$}}}
%\put(0,1000){\makebox(100,100)[l]{{\large $\sim$ $\sqrt{\sigma_{tot}}$}}}
\end{picture}
\end{minipage}
\end{figure}
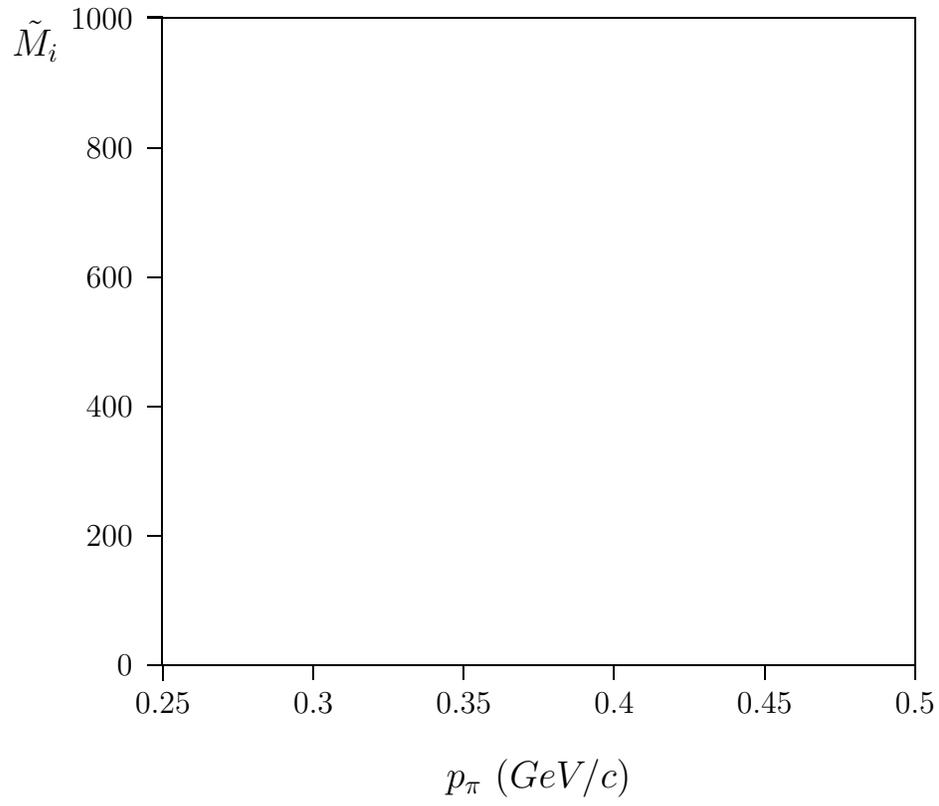

\newpage

\addtocounter{figure}{-1}
\begin{figure}[t]
\caption{e)~The $\pi^+\pi^+n$\,--\,channel. \protect \newline
Compared are the fits with the full data set and with a
partial data set
(lower and
upper line, respectively, 4 points out of 20 are suppressed).
}
\begin{minipage}{100mm}
\textheight 5mm
\textwidth 2mm
\setlength{\unitlength}{0.1mm}
\begin{picture}(1000,1332)
  \put(300,300){\framebox(1000,860)}
  \put(150,200){\makebox(300,100){0.25}}
  \put(300,299){\line(0,-1){20}}
  \put(350,200){\makebox(300,100){0.3}}
  \put(500,300){\line(0,-1){20}}
  \put(550,200){\makebox(300,100){0.35}}
  \put(700,300){\line(0,-1){20}}
  \put(750,200){\makebox(300,100){0.4}}
  \put(900,300){\line(0,-1){20}}
  \put(950,200){\makebox(300,100){0.45}}
  \put(1100,300){\line(0,-1){20}}
  \put(1150,200){\makebox(300,100){0.5}}
  \put(1300,300){\line(0,-1){20}}
  \put(-40,250){\makebox(300,100)[r]{   0}}
  \put(300,300){\line(-1,0){20}}
  \put(-40,422){\makebox(300,100)[r]{ 200}}
  \put(300,472){\line(-1,0){20}}
  \put(-40,594){\makebox(300,100)[r]{ 400}}
  \put(300,644){\line(-1,0){20}}
  \put(-40,766){\makebox(300,100)[r]{ 600}}
  \put(300,816){\line(-1,0){20}}
  \put(-40,938){\makebox(300,100)[r]{ 800}}
  \put(300,988){\line(-1,0){20}}
  \put(-40,1110){\makebox(300,100)[r]{1000}}
  \put(300,1162){\line(-1,0){20}}
\put(300,50){\makebox(1000,200){{\large $p_{\pi}\,\,(GeV/c)$}}}
\put(100,1080){\makebox(100,100)[l]{{\large $\tilde M_i$}}}
%\put(0,1000){\makebox(100,100)[l]{{\large $\sim$ $\sqrt{\sigma_{tot}}$}}}
\end{picture}
\end{minipage}
\end{figure}
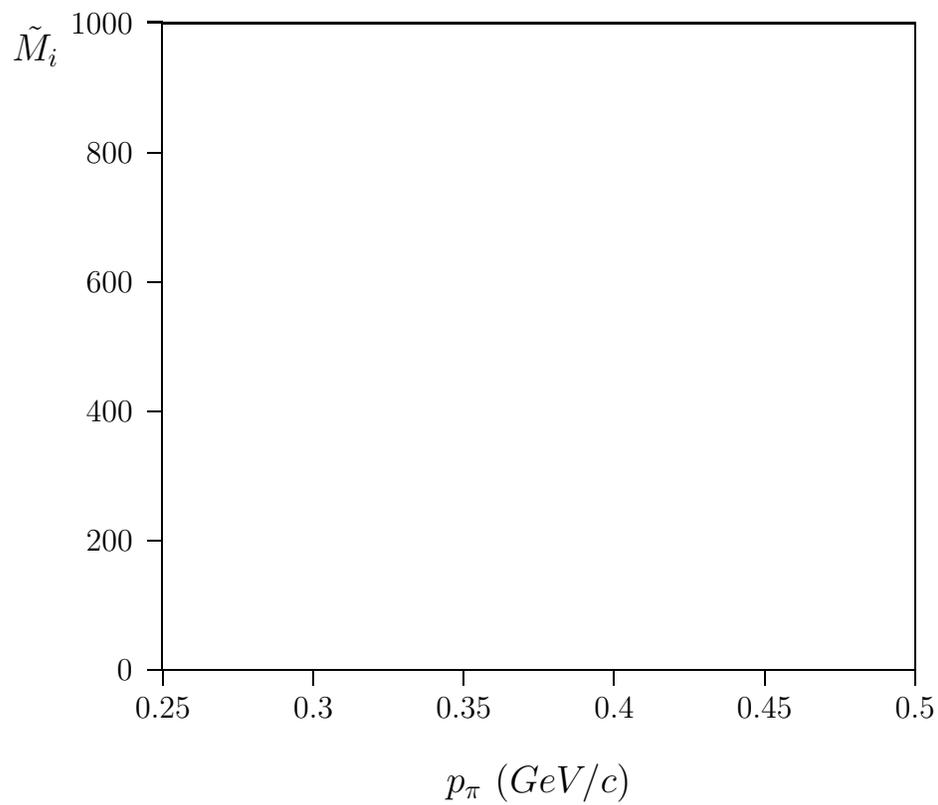

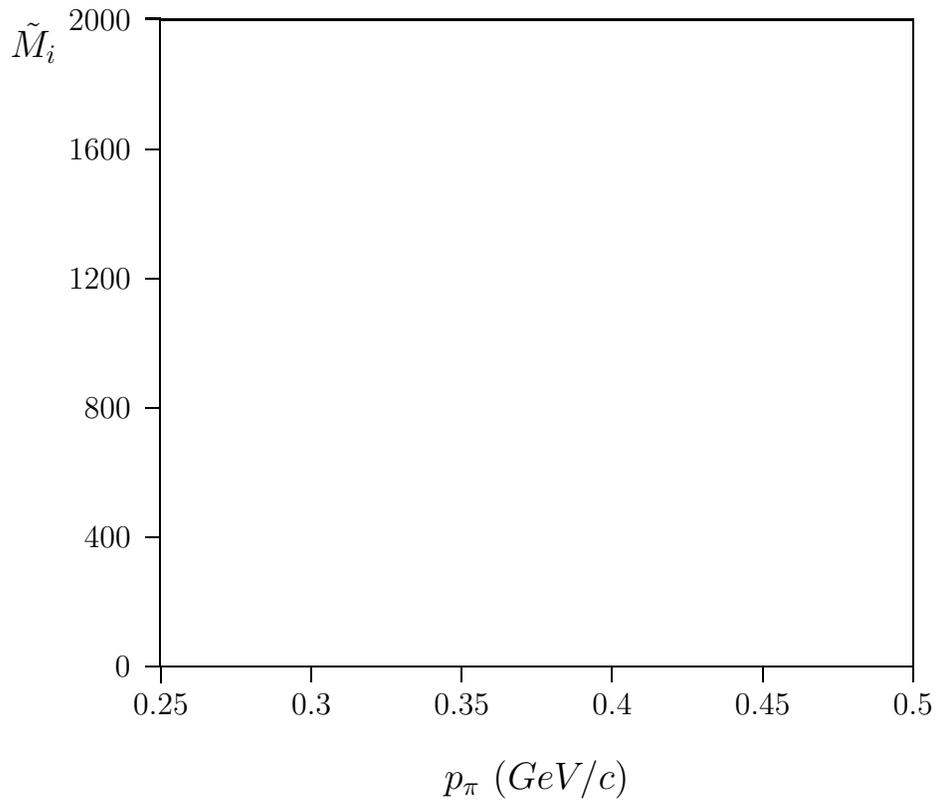
\end{document}